\documentclass[journal]{IEEEtran}
\usepackage{amsmath,amsfonts}
\usepackage{algorithmic}
\usepackage{algorithm}
\usepackage{array}
\usepackage[caption=false,font=footnotesize,labelfont=rm,textfont=rm]{subfig}
\usepackage{textcomp}
\usepackage{stfloats}
\usepackage{url}
\usepackage{verbatim}
\usepackage{graphicx}
\usepackage{cite}
\usepackage{booktabs}
\usepackage{multirow}
\usepackage{amssymb}
\usepackage{tabularx}
\usepackage[colorlinks=true, linkcolor=blue, citecolor=blue, urlcolor=blue]{hyperref}
\newcolumntype{Y}{>{\centering\arraybackslash}X}
\usepackage{makecell}

\begin{document}

\title{Low-Cost Architecture and Efficient Pattern Synthesis for Polarimetric Phased Array Based on Polarization Coding Reconfigurable Elements}

\author{Yiqing Wang, Jian Zhou, Chen Pang, Wenyang Man, Zixiang Xiong, Ke Meng, \\Zhanling Wang, Yongzhen Li
\thanks{This work was supported by the National Natural Science Foundation of China Grant 62301580, Grant 61921001. \emph{(Corresponding author: Chen~Pang.)}}
\thanks{The authors are with the State Key Laboratory of Complex Electromagnetic Environment Effects on Electronics and Information System, College of Electronic Science and Technology, National University of Defense Technology, Changsha 410073, China (email: wangyiqing@nudt.edu.cn, zhoujianlfx@163.com, pangchen1017@hotmail.com, wenyang622@outlook.com, xiongzixiang23@nudt.edu.cn, mengke@nudt.edu.cn, zlwangnudt@outlook.com, e0061@sina.com).}}

\markboth{IEEE TRANSACTIONS ON ANTENNAS AND PROPAGATION,~Vol.~XX, No.~XX, XX}%
{Shell \MakeLowercase{\textit{et al.}}: A Sample Article Using IEEEtran.cls for IEEE Journals}

\IEEEpubid{0000--0000/00\$00.00~\copyright~2021 IEEE}

\maketitle

\begin{abstract}
Polarimetric phased arrays (PPAs) enhance radar target detection and anti-jamming capabilities, but their conventional dual transmit/receive (T/R) channel architecture leads to high cost and system complexity. To address these limitations, this paper proposes a polarization-coding reconfigurable phased array (PCRPA) and associated pattern synthesis techniques, which reduce the channel count while preserving key performance. In the PCRPA, each antenna element connects to a single T/R channel and is equipped with a two-level RF switch, enabling real-time control of its polarization state and subarray grouping. By optimizing both the element polarization codes and the excitation weights, the array can synthesize arbitrarily polarized and dual-polarized beams. 
Simulation results show that the proposed approach achieves suppressed cross-polarization and comparable sidelobe levels compared to conventional PPAs across a wide scan range, with performance improvements being more pronounced in larger arrays. The inherent channel reduction does, however, incur a trade-off in terms of radiated power and directivity. Experimental validation using an $8\times 8$ X-band array antenna confirms the feasibility and effectiveness of the proposed system. The PCRPA architecture and the accompanying synthesis methods offer a cost-effective solution for large-scale PPA systems, maintaining sidelobe and polarization control with significantly reduced hardware complexity.
\end{abstract}


\section{Introduction}
\IEEEPARstart{P}{olarimetric} phased arrays (PPA) tend to be adopted by many radar systems to enhance their measurement performance within complex environments. The extra dimension of polarization can provide detailed information to detect targets in the presence of strong clutter\cite{hurtado2008polar,li2014ground,cheng2015antenna,wu2024a,an2024ground}, interference\cite{compton1981on,lu2021adaptive,yin2022radio,pan2023joint,an2024radio,xie2025coarray}, or other adjacent objects\cite{zhu2024principles,zhu2024a,liu2024robust}. These applications rely on the beamforming capabilities of dual- or arbitrarily polarized patterns, which require independent channels of orthogonal polarization states, usually horizontal (H) and vertical (V). Therefore, conventional PPA architectures generally consist of dual-polarized antenna elements with transmit/receive (T/R) channels connected to each of their ports\cite{salazar2015transmit,ta2015crossed,salazar2015tr,mirmozafari2019dual}. Compared with single-polarized phased arrays of the same size, the number of channels in the T/R modules of PPAs is doubled.

Conventional architectures offer high performance but significantly increase the manufacturing cost and complexity of PPA systems, especially for large-scale arrays. A typical cost distribution for phased arrays given by \cite{herd2016the} shows that the T/R modules can account for half the system cost, whose components grow proportionally with the number of elements. Meanwhile, many recently reported advanced PPA radar systems are on the scale of 1000-element or even bigger\cite{fulton2017cylindrical,zhang2022cylindrical,palmer2023horus} to obtain high space resolution and more degrees of freedom (DoFs) for signal processing. Therefore, for these large-scale PPA systems, the heavy use of T/R channels under conventional architectures will make the total cost and complexity fairly high, which also reveals the practical opportunity for cost reduction.

Over the past few decades, techniques like sparse arrays, thinned arrays, and subarrays have been thoroughly studied. They can significantly reduce the cost and complexity of phased arrays and directly apply to PPAs\cite{rocca2016uncon,mailloux2009irreg}. However, these array designs are relatively fixed and static, so the configuration must be determined before manufacturing and cannot be changed. This will limit some crucial performance of radar systems, especially the power and scan range. In addition, polarization reconfigurable antennas are generally designed with single ports, requiring half the T/R channels compared to dual-polarized antennas\cite{nishamol2011an,lin2017multi,chen2021a}. If they are employed as PPA elements, the system cost and complexity will also be significantly reduced. Novel algorithms considering element polarization as variables can be proposed to ensure the performance of polarization pattern synthesis during the working process. Since the polarization states are adjusted in real-time, the design can be regarded as a dynamic thinned array with polarization as a new DoF. 

\IEEEpubidadjcol
Current research related to this issue can be divided into two schemes. The first has thoroughly investigated the application of polarization-reconfigurable elements in metasurfaces\cite{yang2021recon,yang2023a}, transmitarrays\cite{wan2023code}, and other non-electronically scanned array antennas\cite{chen2022,wang2022}. These works are instructive, but their array architectures and pattern synthesis models differ from the PPAs. The second scheme explores the idea of mixed polarized elements in PPA pattern synthesis but does not involve specific architecture design or beamforming requirements\cite{barbetty2012,salazar2024,salazar2024a,wang2024low-cost,zhou2024}. For example, in \cite{salazar2024,salazar2024a}, a signal processing method called cross-polarization canceller (XPC) uses a selection of cross-polarized elements to improve the cross-polarization level (XPL) of beam patterns by canceling the corresponding field components. However, it is based on the conventional architecture and requires the phased array to be fully digital. Several architecture models have been reported in \cite{wang2024low-cost,zhou2024}, but they do not cover the commonly used modes of PPA radars, such as alternately transmitting and simultaneously receiving (ATSR) and simultaneously transmitting and simultaneously receiving (STSR).

This paper proposes an architecture and corresponding polarization pattern synthesis methods for PPA radars to reduce the cost and complexity while ensuring beamforming performance. The proposed architecture employs elements with polarization reconfigurable ability. Given that the polarization states of each element are limited and independently controllable, this architecture is termed the polarization coding reconfigurable phased array (PCRPA). In the hardware design, each T/R channel has two switches: the antenna-connected switch for the antenna polarization state selection and the transceiver-connected switch for the subarray selection. This paper proposes novel constraints, derived from antenna radiation and array geometry analysis, to optimize pattern performance and enhance algorithm efficiency. Combined with classical optimization algorithms, efficient synthesis methods for the polarization pattern of the PCRPA are proposed. The main contributions of this paper are listed below. 
\begin{enumerate}
    \item A novel array architecture considering multiple applications for PPA radars is proposed. The designed architecture enables the real-time coding of polarization states and subarrays for reconfigurable antenna elements, allowing for the synthesis of arbitrarily polarized and dual-polarized beam patterns. Compared to conventional architectures, the proposed design offers a cost-effective advantage by reducing the number of active channels in the T/R modules by half. 
    \item The pattern performance of arbitrarily polarized and dual-polarized beams is analyzed through simulations. Two pattern synthesis methods are proposed based on the different performance to efficiently control power and polarization properties, where elements' polarization states are introduced as variables.
    \item The effectiveness of the proposed architecture and pattern synthesis methods is verified through electromagnetic simulations and experiments. A detailed discussion is provided on the impact of the desired polarization state, beam direction, and array size. The discussion points out that the array size is a crucial factor for the performance of the proposed design.
\end{enumerate}

The remainder of this paper is organized as follows. Section \ref{sec2} details the architecture design of the proposed PCRPA. Section \ref{sec3} presents the theoretical foundation, formulating the efficient pattern synthesis methods. Section \ref{sec4} provides a comprehensive simulation validation and performance analysis, including parameter sensitivity studies. Section \ref{sec5} offers experimental verification on a physical array prototype, discussing the agreement between measured and simulated results. Finally, Section \ref{sec6} concludes the paper.

Notations: Some important symbols and notational definitions used in this article are given in Table \ref{tab_not}. In this paper, boldface upper-case letters denote matrices, boldface lower-case letters denote column vectors, and standard lowercase letters denote scalars.

\begin{table}[htbp]
\centering
\caption{Notation}
\label{tab_not}
\setlength{\extrarowheight}{0.25\baselineskip}
\begin{tabularx}{0.9\columnwidth}{>{\centering\arraybackslash}X >{\raggedright\arraybackslash}X}
\hline
\hline
\textbf{Symbol} & \textbf{Description} \\
\midrule
\( \theta \) & Elevation angle \\
\( \phi \) & Azimuth angle \\
\( \gamma \) & Auxiliary polarization angle \\
\( \eta \) & Polarization phase angle \\
 H & Horizontal \\
 V & Vertical \\
 CoP & Co-polarization \\
 XP & Cross-polarization\\
 PPA  & Polarimetric phased array \\
 CPPA & Conventional polarimetric phased array \\
 PCRPA & Polarization coding reconfigurable phased array \\
 PSL & Peak sidelobe level \\
 XPL & Cross-polarization level \\
 AEP & Active element pattern \\
\hline
\hline
\end{tabularx}
\end{table}

\section{General Architecture of The PCRPA}\label{sec2}

\begin{figure*}[htbp]
\centering
\subfloat[]{\includegraphics[width=16cm]{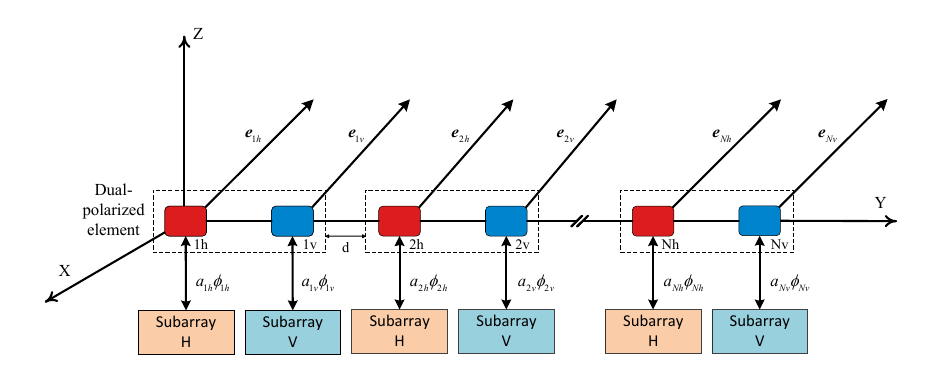}%
\label{fig_sch_conv}}
\\
\subfloat[]{\includegraphics[width=9cm]{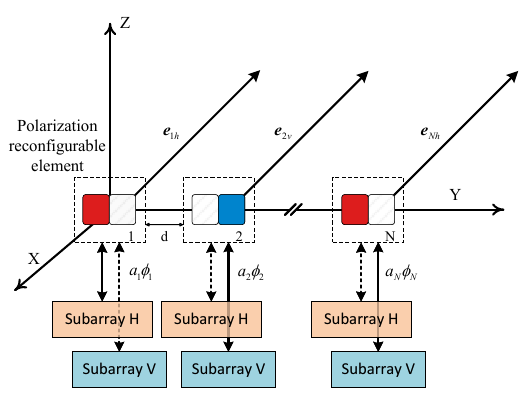}%
\label{fig_sch_prop}}
\hfil
\subfloat[]{\includegraphics[width=9cm]{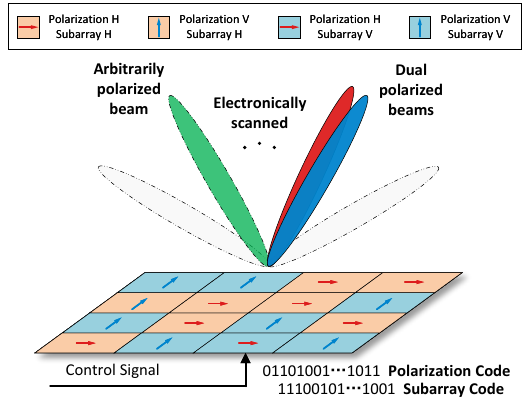}%
\label{fig_sch_lay}}
\caption{Illustration of the conventional and the proposed PPA architectures. (a) Conventional architecture. (b) Proposed architecture. (c) Schematic diagram of the proposed architecture, which can produce dual-polarized or arbitrarily polarized beams by controlling the polarization states and subarrays of elements.}
\label{fig_sch}
\end{figure*}

This section begins with an overview of the conventional PPA architecture. As illustrated in Fig. \ref{fig_sch_conv}, the architecture consists of $N$ elements spaced at a uniform distance $d$ on the YOZ plane. In this architecture, the H and V polarization ports of each dual-polarized antenna element are connected to independent T/R channels (demonstrated by weight $a_{ih}\phi_{ih}$ or $a_{iv}\phi_{iv}$) and are fixedly assigned to corresponding H-polarized or V-polarized subarrays. The system synthesizes beams independently within the H and V subarrays via separate beamforming networks, which are then coherently combined in the far-field to achieve dual-polarized beam calibration or to synthesize beams with arbitrary polarization states. In this case, the electric field of beams formed by H and V subarrays can be expressed as:
\begin{equation}
    \left\{ {\begin{array}{l}
    {\boldsymbol{E}_h\left( {\theta ,\phi } \right) = \sum_{i=1}^{N} \boldsymbol{e}_{ih}\left( {\theta ,\phi } \right)}\\
    {\boldsymbol{E}_v\left( {\theta ,\phi } \right) = \sum_{i=1}^{N} \boldsymbol{e}_{iv}\left( {\theta ,\phi } \right)}
\end{array}} \right.
\label{eq_EhEv}
\end{equation}
where $\boldsymbol{e}_{ih}(\theta,\phi)$ and $\boldsymbol{e}_{iv}(\theta,\phi)$ denote the electric field vectors of the $i$-th H and V ports, and the effects of amplitude weights $a_{ih}$, $a_{iv}$ and phase weights $\phi_{ih}$, $\phi_{iv}$ have already been included. However, this conventional approach requires a large number of TR channels, leading to high system cost and complexity.

To address this issue, this paper proposes a PCRPA architecture, as illustrated in Fig. \ref{fig_sch_prop}, where the number and positions of elements remain consistent with those in Fig. \ref{fig_sch_conv}. The idea for cost reduction is achieved through the resource sharing of T/R channels among the polarization ports and subarray modules. Specifically, polarization-reconfigurable antenna elements are employed, allowing the dual-polarized ports of each element to time-share a single T/R channel, thereby halving the total number of channels. Furthermore, to address the beam reconfiguration requirements for dual-polarized pattern synthesis, a similar approach is applied, enabling the H- and V-polarized subarrays to also time-share a T/R channel. These two levels of reconfiguration collectively form an array architecture in which both the polarization and subarray connections are dynamically reconfigurable.

In this way, the PCRPA architecture realizes beam synthesis through a dual-coding mechanism. First, subarray coding determines the set of element indices, denoted as $\boldsymbol{I}_j$, excited for the $j$-th beam. Second, polarization coding defines the sets of indices for all elements in the array excited through their H ports and V ports, denoted as $\boldsymbol{I}_h$ and $\boldsymbol{I}_v$. The total radiation pattern of beam $j$ is the superposition of the fields radiated by the elements in the intersection set of $\boldsymbol{I}_{jh}$ and $\boldsymbol{I}_{jv}$, defined as:
\begin{equation}
    \left\{ {\begin{array}{l}
    {\boldsymbol{I}_{jh} = \boldsymbol{I}_j \cap \boldsymbol{I}_h}\\
    {\boldsymbol{I}_{jv} = \boldsymbol{I}_j \cap \boldsymbol{I}_v}
\end{array}} \right.
\label{eq_IjhIjv}
\end{equation} 
And the electric field of the $j$-th beam can be expressed as:
\begin{equation}
    \boldsymbol{E}_{j}(\theta,\phi) = 
    \sum_{i=1}^{N_{jh}}\boldsymbol{e}_{I^{jh}_{i}h}(\theta,\phi) + \sum_{i=1}^{N_{jv}}\boldsymbol{e}_{I^{jv}_{i}v}(\theta,\phi)
    \label{eq_Ej}
\end{equation}
where $N_{jh}$ and $N_{jv}$ are the number of elements in sets $\boldsymbol{I}_{jh}$ and $\boldsymbol{I}_{jv}$. Here, vectors $\boldsymbol{I}^{jh}$ and $\boldsymbol{I}^{jv}$ should satisfy the constraints followed:
\begin{equation}
    \left\{ {\begin{array}{l}
    {1\leq I^{jA}_{i}\leq N,\:
    I^{jA}_{i}\in \mathbb{Z},i=1,2,\cdots,N_{jA}}\\
    {I^{jA}_{m}\neq I^{jA}_{n},\:
    \forall m,n\in \{1,2,\cdots,N_{jA}\},m\neq n}\\
    {\boldsymbol{I}^{jh} \cap \boldsymbol{I}^{jv} = \varnothing}
\end{array}} \right.
\label{eq_Ijconst}
\end{equation}
where $A\in\{ h,v\}$ represents the polarization port.

Based on Equation \ref{eq_Ej}, the proposed PCRPA architecture is able to form electrially scanned beams of either arbitrary or dual polarization, shown in Fig. \ref{fig_sch_lay}. This is owing to two key aspects: Firstly, the code vectors $\boldsymbol{I}_{jh}$ and $\boldsymbol{I}_{jv}$ create irregular distributions of phase centers across the aperture, analogous to the synthesis principles of sparse arrays, which enables the formation of high-gain, low-sidelobe beams. Secondly, by dynamically configuring the number and excitation coefficients (amplitude and phase) of elements fed through different polarization ports, the architecture can achieve functionality similar to cross-polarization calibration techniques, allowing for the precise synthesis of any desired polarization state.

\begin{figure*}[htbp]
\centering
\subfloat[]{\includegraphics[width=16cm]{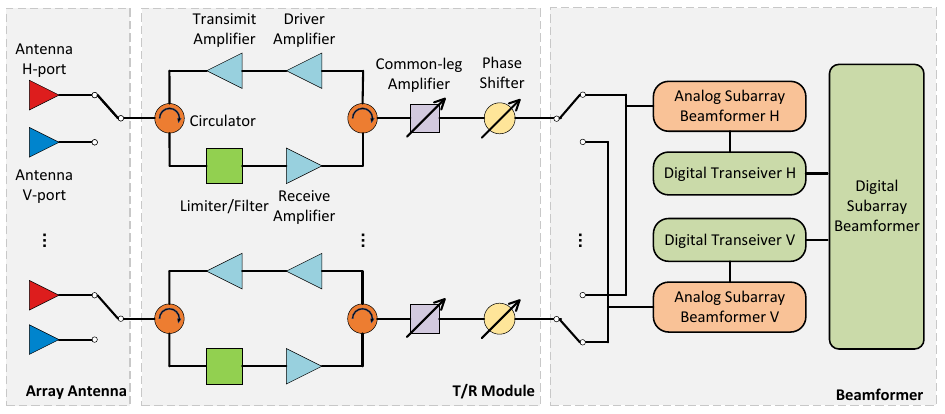}%
\label{fig_bf_analog}}
\\
\subfloat[]{\includegraphics[width=16cm]{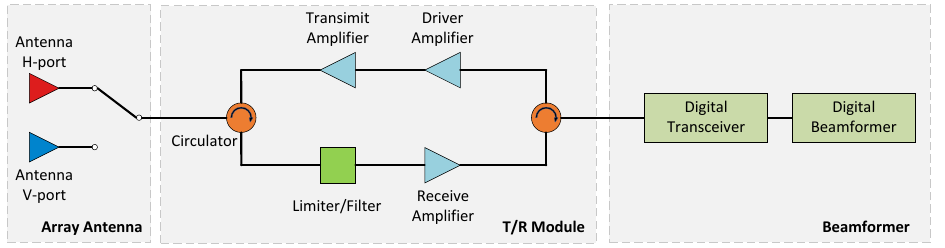}%
\label{fig_bf_digital}}
\caption{The beamformer design of the proposed phased array architecture for different phased array architectures. (a) For subarray-level digital phased array. (b) For fully digital phased array.}
\label{fig_bf}
\end{figure*}

\begin{table*}[htbp]
\centering
\caption{Comparison of PPA Architectures}
\label{tab_archcomp}
\renewcommand{\arraystretch}{1.5}
\begin{tabularx}{2\columnwidth}{>{\raggedright\arraybackslash}p{0.12\linewidth} *{5}{>{\centering\arraybackslash}m{0.15\linewidth}}}
\toprule
\textbf{Architecture} & \textbf{Channel Number} & \textbf{Element Polarization} & \textbf{Beamforming DoFs} & \textbf{Arbitrarily Polarized Beamforming} & \textbf{Dual Polarized Beamforming} \\
\midrule
\textbf{Conventional} & $2N$ & Fixed & Weight & Yes & Yes \\
\addlinespace
\textbf{Ref. \cite{zhou2024}} & $N$ & Fixed & Weight & Yes & No \\
\addlinespace
\textbf{Proposed} & $N$ & Reconfigurable & \makecell{Weight, Port, \\ Subarray} & Yes & Yes \\
\bottomrule
\end{tabularx}
\end{table*}

In terms of the hardware implementation, the two-level coding process of the PCRPA can be realized by the two-level of RF switches shown in Fig. \ref{fig_bf_analog}. Here, RF switches are utilized to connect the array antenna, T/R modules, and beamformers sequentially. Here, the combination of dual-polarized antennas and RF switches may also be replaced by polarization reconfigurable antennas with single ports. However, the latter may not tolerate high transmitting power of radar systems at presence due to components like PIN diodes. Besides, although this paper focuses on the subarray-level digital systems, the PCRPA is also applicable to fully digital phased arrays. In a fully digital architecture shown by Fig \ref{fig_bf_digital}, the subarray coding function can be implemented directly by the digital beamformer\cite{fulton2016digital}, allowing the second-level RF switch to be omitted, thereby simplifying the hardware structure.

To summarize, a comparative analysis is provided by Table \ref{tab_archcomp}. In comparison to the conventional PPAs, both the proposed PCRPA architecture and a previously reported structure by the team \cite{zhou2024} can reduce the number of T/R channels by half, lowering the system cost and complexity. Relative to the prior work, the key difference of the PCRPA lies in beamforming DoFs and capabilities. While \cite{zhou2024} utilized fixed phase shifters to decide element polarization, this work adopted a real-time reconfigurable design. This makes the port polarization a new DoF for beamforming in the proposed architecture. Additionally, the subarray coding DoF is introduced. These enhancements enable the PCRPA to achieve dual-polarized beamforming. Meanwhile, the real-time reconfigurability of polarization coding also leads to better radiation performance, which will be demonstrated later.

\section{Pattern Synthesis for The PCRPA}\label{sec3}

\subsection{Problem Formulation}

Consider a conventional polarimetric phased array antenna with $N$ elements, the vector of its active element patterns (AEPs) can be expressed as:
\begin{equation}
    \boldsymbol{g}_{AB}(\theta,\phi)=\left[ g_{1,AB}(\theta,\phi),...,g_{N,AB}(\theta,\phi) \right]^{T}
    \label{eq_g}
\end{equation}
where $A,B\in\{h,v\}$ represent the polarization of the electric field component and the antenna port, respectively.

The beam patterns are formed for the active phased arrays by adding each AEP with independent weights. In this case, the array pattern can be expressed as\cite{pozar1994the}:
\begin{equation}
    \left\{ {\begin{array}{l}
    {{f_h}\left( {\theta ,\phi } \right) = \boldsymbol{w}_{h}^{T}\boldsymbol{g}_{hh}(\theta,\phi)+\boldsymbol{w}_{v}^{T}\boldsymbol{g}_{hv}(\theta,\phi)}\\
    {{f_v}\left( {\theta ,\phi } \right) = \boldsymbol{w}_{h}^{T}\boldsymbol{g}_{vh}(\theta,\phi)+\boldsymbol{w}_{v}^{T}\boldsymbol{g}_{vv}(\theta,\phi)}
\end{array}} \right.
\label{eq_fwg}
\end{equation}
where $\boldsymbol{w}_{h}$ and $\boldsymbol{w}_{v}$ denote the weight vectors for H and V ports of the array. These vectors can be generally adjusted freely, depending on the amplitudes and phases, to form desired beam patterns. Due to space limitations, this paper only addresses the transmitting problem, which allows phase-only synthesis model below:
\begin{equation}
    \boldsymbol{w}_{h}=\boldsymbol{w}_{v}=\text{exp}(jk\boldsymbol{P}^{T}\boldsymbol{r}_{0})
\label{eq_wPr}
\end{equation}
where $k=2\pi/\lambda$, $\boldsymbol{r}_{0}=\left[ \sin{\theta_{0}}\cos{\phi_{0}},\sin{\theta_{0}}\sin{\phi_{0}},\cos{\theta_{0}} \right]^{T}$, and $\boldsymbol{P}$ represents the position of elements:
\begin{equation}
    \boldsymbol{P}=\left[
    \begin{array}{*{20}{c}}
        p_{1x}&p_{2x}&\cdots&p_{Nx} \\
        p_{1y}&p_{2y}&\cdots&p_{Ny} \\
        p_{1z}&p_{2z}&\cdots&p_{Nz}
    \end{array}
    \right]
\label{eq_P}
\end{equation}

According to the architecture shown in Fig. \ref{fig_sch_prop}, the synthesis of PCRPA patterns depends on variables of the weight and code vectors for antenna elements. The coding process can be expressed by the polarization coding vectors $\boldsymbol{x}_{h}$ and $\boldsymbol{x}_{v}$. The vectors should satisfy:
\begin{equation}
    \left\{ {\begin{array}{l}
    {x_{h},x_{v}\in\{0,1\}}\\
    {\boldsymbol{x}_{h}+\boldsymbol{x}_{v} \leq \boldsymbol{1}}
\end{array}} \right.
\label{eq_x}
\end{equation}

For clear clarification, define code matrix as below:
\begin{equation}
    \left\{ {\begin{array}{l}
    {\boldsymbol{X}_{h}=\text{diag}(\boldsymbol{x}_{h})}\\
    {\boldsymbol{X}_{v}=\text{diag}(\boldsymbol{x}_{v})}
\end{array}} \right.
\label{eq_Xx}
\end{equation}

For any beam synthesized by the PCRPA, its pattern can be then expressed as:
\begin{equation}
    \left\{ {\begin{array}{l}
    {{f_h}\left( {\theta ,\phi } \right) = \boldsymbol{w}_{h}^{T}\boldsymbol{X}_{h}    \boldsymbol{g}_{hh}(\theta,\phi)+\boldsymbol{w}_{v}^{T}\boldsymbol{X}_{v}\boldsymbol{g}_{hv}(\theta,\phi)}\\
    {{f_v}\left( {\theta ,\phi } \right) = \boldsymbol{w}_{h}^{T}\boldsymbol{X}_{h}\boldsymbol{g}_{vh}(\theta,\phi)+\boldsymbol{w}_{v}^{T}\boldsymbol{X}_{v}\boldsymbol{g}_{vv}(\theta,\phi)}
\end{array}} \right.
\label{eq_fwxg}
\end{equation}

\subsection{Arbitrarily Polarized Pattern Synthesis}\label{subsec:arb}

\begin{figure*}[htbp]
\centering
\includegraphics[width=17cm]{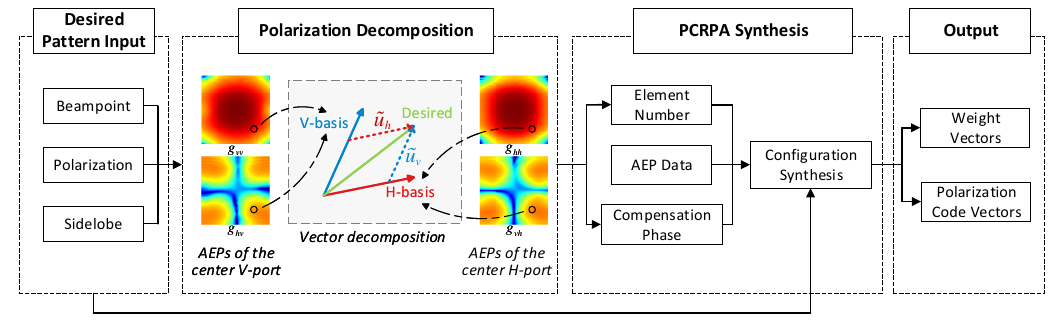}
\caption{Flowchart of the arbitrarily polarized pattern synthesis method.}
\label{fig_flow_single}
\end{figure*}

The subsequent part introduces the arbitrarily polarized pattern synthesis method procedure, with a detailed flowchart illustrated in Fig. \ref{fig_flow_single}.
For arbitrarily polarized pattern synthesis problems, define the desired polarization basis according to Ludwig's polarization definition \uppercase\expandafter{\romannumeral2}\cite{ludwig1973the}:
\begin{equation}
    \left\{ 
    \begin{array}{l}
    {\boldsymbol{e}_{co}=\left[\cos{\gamma,\sin{\gamma}e^{j\eta}}\right]^{T}}\\
    {\boldsymbol{e}_{cr}=\left[-\sin{\gamma,\cos{\gamma}e^{-j\eta}}\right]^{T}}\\
    {\boldsymbol{e}_{co}^{H}\boldsymbol{e}_{cr}=0}
    \end{array}
    \right.
\label{eq_ecoecr}
\end{equation}
where $\gamma\in\left[0^{\circ},90^{\circ}\right]$ and $\eta\in\left[-180^{\circ},180^{\circ}\right)$ represent the parameters of the auxiliary polarization angle and the polarization phase difference. The vectors of $\boldsymbol{e}_{co}$ and $\boldsymbol{{e}_{cr}}$ denote the unit electric field vectors of the co-polarization and cross-polarization components.

According to the conservation of electric fields, the following equation holds:
\begin{equation}
    \left[ \boldsymbol{e}_{co},\boldsymbol{e}_{cr} \right]
    \left[ \begin{array}{l} {f_{co}(\theta,\phi)}\\{f_{cr}(\theta,\phi)} \end{array} \right] = 
    \left[ \boldsymbol{e}_{h},\boldsymbol{e}_{v} \right]
    \left[ \begin{array}{l} {f_{h}(\theta,\phi)}\\{f_{v}(\theta,\phi)} \end{array} \right]
\label{eq_cocr_hv}
\end{equation}
where $\boldsymbol{e}_{h}=\left[1,0\right]$ and $\boldsymbol{e}_{v}=\left[0,1\right]$ represent the unit vectors of the H and V polarization fields.

Let the co-polarization component $f_{co}(\theta,\phi)=1$ and cross-polarization component $f_{cr}(\theta,\phi)=0$, according to Equation (\ref{eq_cocr_hv}), the vector decomposition results of the desired polarization state under the standard H and V basis can be defined as:
\begin{equation}
\begin{aligned}
    \left[ \begin{array}{l} {u_{h}}\\{u_{v}} \end{array} \right] 
    &\triangleq \left[ \begin{array}{l} {f_{h}(\theta,\phi)}\\{f_{v}(\theta,\phi)} \end{array} \right]\\
    &= \left[ \boldsymbol{e}_{h},\boldsymbol{e}_{v} \right]^{-1}\boldsymbol{e}_{co}\\
    &= \left[ \begin{array}{l} {\cos(\gamma)}\\{\sin(\gamma)e^{j\eta}} \end{array} \right]
\end{aligned}
\label{eq_fcofcr}
\end{equation}

Based on Equation (\ref{eq_fcofcr}), the conventional PPAs can synthesize patterns of arbitrarily polarized beams by directly multiplying the corresponding weights $\cos{\gamma}$ and $\sin{\gamma}e^{-j\eta}$ into the polarization channels H and V\cite{zhou2024}. However, this method ignores the cross-polarization components radiated at the H and V ports. If they are considered, there will be:
\begin{equation}
    \begin{aligned}
    f_{cr}(\theta,\phi) = 
    &\sin(\gamma)\cos(\gamma)\left[-f_{hh}(\theta,\phi)+f_{vv}(\theta,\phi)\right]\\
    &+\cos^2(\gamma)e^{-j\eta}f_{vh}(\theta,\phi)
    -\sin^2(\gamma)e^{j\eta}f_{hv}(\theta,\phi)
    \end{aligned}
\label{eq_fcrneq0}
\end{equation}
Even if $f_{hh}(\theta,\phi)$ and $f_{vv}(\theta,\phi)$ are highly matched, due to the influence of the latter two terms, the cross-polarization component of the beam $f_{co}(\theta,\phi)$ is difficult to suppress. Especially when the scanning angle increases, the polarization purity of the beam will decrease rapidly.

In this case, this paper proposes a decomposition model based on antenna AEPs. The idea is shown in the second part of Fig. \ref{fig_flow_single}. The green arrow represents the desired polarization state of the beam pattern. It can be decomposed according to the polarization basis represented by the red and blue arrows. The vectors of the basis are defined by the AEPs of an embedded element taken from the array center:
\begin{equation}
    \left\{ {\begin{array}{l}
    {\boldsymbol{\Tilde{e}}_{h}=\boldsymbol{g}_{h}(\theta_0,\phi_0)/\left\| \boldsymbol{g}_{h}(\theta_0,\phi_0) \right\|}\\
    {\boldsymbol{\Tilde{e}}_{v}=\boldsymbol{g}_{v}(\theta_0,\phi_0)/\left\| \boldsymbol{g}_{v}(\theta_0,\phi_0) \right\|}
\end{array}} \right.
\label{eq_ehpevp}
\end{equation}
where
\begin{equation}
    \left\{ {\begin{array}{l}
    {\boldsymbol{g}_{h}(\theta_0,\phi_0)={\left[ {\begin{array}{*{20}{c}}
    {{g_{hh}}\left( {{\theta _0},{\phi _0}} \right)}&{{g_{vh}}\left( {{\theta _0},{\phi _0}} \right)}
    \end{array}} \right]^T}}\\
    {\boldsymbol{g}_{v}(\theta_0,\phi_0)=}{\left[ {\begin{array}{*{20}{c}}
    {{g_{hv}}\left( {{\theta _0},{\phi _0}} \right)}&{{g_{vv}}\left( {{\theta _0},{\phi _0}} \right)}
    \end{array}} \right]^T}
\end{array}} \right.
\label{eq_ghgvgmax}
\end{equation}

And the decomposition process can be expressed as:
\begin{equation}
    \left[ \begin{array}{l} {\Tilde{u}_{h}}\\{\Tilde{u}_{v}} \end{array} \right] =
    \left[ \boldsymbol{\Tilde{e}}_{h},\boldsymbol{\Tilde{e}}_{v} \right]^{-1}\boldsymbol{e}_{co}
\label{eq_decomp}
\end{equation}
where $\Tilde{u}_{h}$ and $\Tilde{u}_{v}$ are complex numbers that can measure the contribution of each port to the desired polarization.

Since this paper focuses on the phase-only synthesis, the amplitude of weights $\Tilde{u}_{h}$ and $\Tilde{u}_{v}$ can be used to synthesize the number of elements excited by H and V ports:
\begin{equation}
    \left\{ {\begin{array}{l}
    {n_{h} \triangleq \lVert \boldsymbol{x}_h \rVert_1
    = \lceil {\frac{{\left| \Tilde{u}_{h} \right|}}{{\left| \Tilde{u}_{h} \right| + \left| \Tilde{u}_{v} \right|}}N} \rceil}\\
    {n_{v} \triangleq \lVert \boldsymbol{x}_v \rVert_1
    = N-n_{h}}
\end{array}} \right.
\label{eq_num}
\end{equation}

In addition, a compensation phase is applied to the channels of V ports to ensure the desired polarization, which is defined:
\begin{equation}
  \beta=\angle \left( {\frac{\Tilde{u}_{v}}{\Tilde{u}_{h}}} \right)
\label{eq_compphase}
\end{equation}
This phase component is directly multiplied by the weight vector of $\boldsymbol{w}_{v}$ given by Equation (\ref{eq_wPr}):
\begin{equation}
    \boldsymbol{w}_{v}=\text{exp}\left[j(k\boldsymbol{P}^{T}\boldsymbol{r}_{0}+\beta)\right]
\label{eq_wv}
\end{equation}

\begin{figure}[htbp]
\centering
\includegraphics[width=6cm]{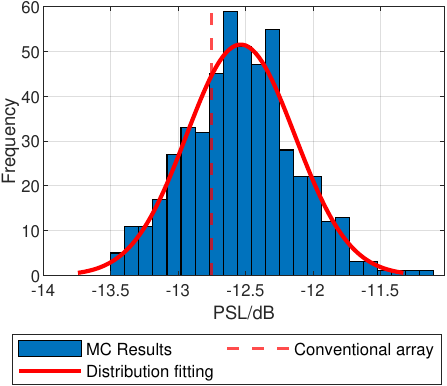}
\caption{Histogram of the Monte Carlo simulation results concerning arbitrarily polarized pattern synthesis.}
\label{fig_mc1}
\end{figure}

The array configuration of $\boldsymbol{x}_{h}$ and $\boldsymbol{x}_{v}$ is yet to be synthesized, which strongly relates to the sidelobe levels of the patterns. To fully investigate the pattern characteristics under the PCRPA model, Fig. \ref{fig_mc1} gives a histogram of the Monte Carlo simulation results concerning this method. The array size is $16 \times 16$, with the desired polarization of V polarization steered at $(120^{\circ},15^{\circ})$. The red dot line marks the PSL of the CPPA under the same circumstances using the weight vector $\boldsymbol{w}_{v}$ given in Equation (\ref{eq_wPr}). The distribution fitting represented by the red solid line shows that 29.74\% of the samples have PSLs lower than the conventional architecture, and the worst case is only 1.63 dB higher. This indicates a relatively concentrated distribution, which is convenient to optimize.

The proposed algorithm, outlined in Algorithm \ref{alg:single}, takes the AEPs of all ports, along with performance specifications (e.g., beam steering, polarization, and sidelobe level), and outputs the polarization code and weight vectors accroding to Fig. \ref{fig_flow_single}. A central challenge addressed by this method is the mitigation of mutual coupling effects between elements and ports. To this end, the algorithm incorporates a two-fold strategy: First, while the polarization basis in Step 2 is constructed using the data of the center embedded element, the array pattern synthesis in Step 5 is computed based on the full AEP model. This model utilizes the full-wave simulated or measured data of all ports, thereby accounting for mutual coupling. Second, to compensate for any residual inaccuracies arising from the single-element estimation in Step 2, Step 7 provides an adjustment mechanism. Specifically, if the pattern requirements are not met upon reaching the maximum generation for a given $n_h$ and $n_v$, the ratio of H-to-V elements is adaptively modified according to a predefined rule (e.g., reducing one H element per adjustment). Furthermore, the random number generator in Step 5, whose distribution is informed by the Monte Carlo simulation results in Fig. \ref{fig_mc1}, can be seamlessly replaced by other optimization algorithms for initial value generation if required.

\begin{algorithm}[H]
\caption{The proposed arbitrarily polarized pattern synthesis method for the PCRPA.}\label{alg:single}
\begin{algorithmic}[1]
\STATE Input the data of active element patterns $\boldsymbol{g}_{AB}(\theta,\phi)$, obtained from either simulations or experiments. Initialize the beam direction $(\theta_0,\phi_0)$, desired polarization state $(\gamma_0,\eta_0)$, the pattern requirements $\Gamma_{PSL}$ and $\Gamma_{XPL}$, and the mainlobe region $\Omega_{mb}$. Set the termination condition of the run time or the maximum number of generations.
\STATE Calculate the decomposition results of the desired polarization state based on the AEP model given by Equation (\ref{eq_ehpevp})-(\ref{eq_compphase}).
\STATE Calculate the number of H and V polarized elements $n_h$, $n_v$, and the compensation phase $\beta$ according to the polarization basis decomposition results.
\STATE Check whether the number of H and V polarized elements is reasonable. If yes, proceed to Step 5; otherwise, stop the synthesis procedure and return the failure prompt.
\STATE Generate the array configuration $\boldsymbol{x}_h$, $\boldsymbol{x}_v$ according to $n_h$, $n_v$ using random number generators. Calculate the patterns $f_{co}(\theta,\phi)$, $f_{cr}(\theta,\phi)$ using Equation (\ref{eq_fwxg}) and (\ref{eq_cocr_hv}).
\STATE Check if the PSL and XPL of the patterns $f_{co}(\theta,\phi)$, $f_{cr}(\theta,\phi)$ meet the requirements of $\Gamma_{PSL}$, $\Gamma_{XPL}$. If yes, exit the procedure and return the synthesis results; otherwise, proceed to Step 7.
\STATE Check if the iteration has reached the maximum generation. If yes, adjust $n_{h}$, $n_{v}$ and repeat Steps 4 to 6; otherwise, repeat Steps 5 to 6.
\end{algorithmic}
\label{alg1}
\end{algorithm}

\subsection{Dual-polarized Pattern Synthesis}
For the measurement applications of the STSR mode, a pair of orthogonally polarized beams is required simultaneously. Therefore, all the elements are split into two groups to synthesize the beams. This can be represented by the subarray coding vectors $\boldsymbol{x}_{1}$ and $\boldsymbol{x}_{2}$, where $x_{1},x_{2}\in \{ 0,1 \}$ and $x_{i,j}=1$ means the $i$-th element belongs to the $j$-th beam. Take the beam $1$ as an example, Equation (\ref{eq_fwxg}) can be transferred to:
\begin{equation}
    \left\{ {\begin{array}{l}
    {{f_{1h}}\left( {\theta ,\phi } \right) = \boldsymbol{w}_{h}^{T}\boldsymbol{X}_{1h}    \boldsymbol{g}_{hh}(\theta,\phi)+\boldsymbol{w}_{v}^{T}\boldsymbol{X}_{1v}\boldsymbol{g}_{hv}(\theta,\phi)}\\
    {{f_{1v}}\left( {\theta ,\phi } \right) = \boldsymbol{w}_{h}^{T}\boldsymbol{X}_{1h}\boldsymbol{g}_{vh}(\theta,\phi)+\boldsymbol{w}_{v}^{T}\boldsymbol{X}_{1v}\boldsymbol{g}_{vv}(\theta,\phi)}
\end{array}} \right.
\label{eq_f1wx1g}
\end{equation}
where
\begin{equation}
    \left\{ {\begin{array}{l}
    {\boldsymbol{X}_{1h}=\text{diag}(\boldsymbol{x}_{1h})}\\
    {\boldsymbol{X}_{1v}=\text{diag}(\boldsymbol{x}_{1v})}\\
    {\boldsymbol{x}_{1h}=\boldsymbol{x}_{1}\odot \boldsymbol{x}_{h}}\\
    {\boldsymbol{x}_{1v}=\boldsymbol{x}_{1}\odot \boldsymbol{x}_{v}}
\end{array}} \right.
\label{eq_Xxx}
\end{equation}

The model given in Equation (\ref{eq_Xxx}) involves nonlinear vector dot products, making direct solution computationally complex. A divide-and-conquer approach is therefore adopted. The problem is decomposed into two sequential steps: first, the subarray coding vectors $\boldsymbol{x}_1$ and $\boldsymbol{x}_1$, which determine the co-polar pattern performance for dual beams, are synthesized. Subsequently, the polarization coding vectors $\boldsymbol{x}_h$ and $\boldsymbol{x}_v$ are optimized using Algorithm \ref{alg:single} described in the previous subsection for arbitrary polarization pattern synthesis, with detailed steps provided later.

\begin{figure}[htbp]
\centering
\includegraphics[width=6cm]{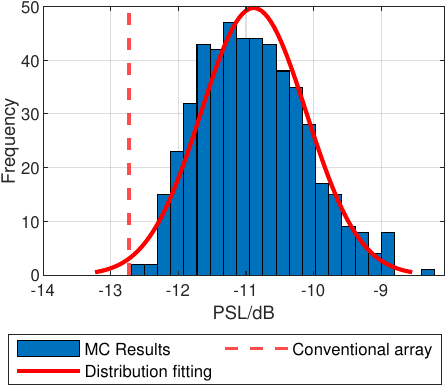}
\caption{Histogram of the Monte Carlo simulation results concerning dual-polarized pattern synthesis.}
\label{fig_mc2}
\end{figure}

A Monte Carlo simulation given by Fig. \ref{fig_mc2} has been carried out to investigate the PSL distribution under the dual-polarized pattern synthesis model. According to the divide-and-conquer approach, assume that $\boldsymbol{x}_1=\boldsymbol{x}_h=\boldsymbol{x}_{1h}$ and $\boldsymbol{x}_2=\boldsymbol{x}_v=\boldsymbol{x}_{2v}$. The beampoint is set at $(120^{\circ},15^{\circ})$, with the PSL taking the maximum of the H and V beams. The red dot line marks the value calculated under the CPPA using Equation (\ref{eq_wPr}). Unlike the phenomenon revealed by Fig. \ref{fig_mc1}, this histogram shows a more dispersed distribution, with the worst PSL 5 dB higher than the CPPA. In addition, the fitting curve marked by the red solid line indicates that only 0.92\% of the configurations have PSLs that are no worse than that of the CPPA.

\begin{figure}[htbp]
\centering
\includegraphics[width=6.5cm]{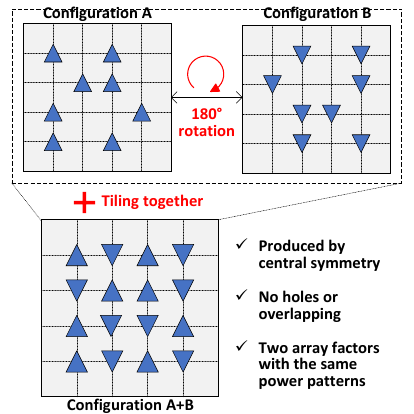}
\caption{Schematic diagram of the central symmetry constraint.}
\label{fig_cs}
\end{figure}

Therefore, an optimized model with a central symmetry constraint is introduced to derive low PSL results. This model can be solved using binary genetic algorithm (BGA) methods and has been discussed in detail in previous work by the authors\cite{wang2024low-cost}. The central symmetry constraint means that two configurations forming the H and V beams are centrally symmetric shown in Fig. \ref{fig_cs}. In this case, the array factors produced by these configurations will have the same power patterns, which have been theoretically demonstrated in \cite{wang2024low-cost}.

This property ensures the optimization model derives highly matched co-polar beam patterns. The constraint can be expressed by the following equation:
\begin{equation}
    \boldsymbol{X}_{2}=\boldsymbol{1}-\text{rot180}(\boldsymbol{X}_{1})
\label{eq_rot180}
\end{equation}
where $\text{rot180(}\cdot\text{)}$ means to rotate a matrix 180 degrees.

From the perspective of matrix operations, Equation (\ref{eq_rot180}) can be further represented in vector form using the subarray code vectors $\boldsymbol{x}_{1}$ and $\boldsymbol{x}_{2}$:
\begin{equation}
    \left \{
    \begin{aligned}
        & \boldsymbol{x}_{1} = 
        \left[
        \begin{array}{*{20}{c}}
            \boldsymbol{x}_{u} \\
            \boldsymbol{1}-\boldsymbol{T}\boldsymbol{x}_{u}
        \end{array}
        \right]\\
        & \boldsymbol{x}_{2} = 
        \left[
        \begin{array}{*{20}{c}}
            \boldsymbol{1}-\boldsymbol{x}_{u} \\
            \boldsymbol{T}\boldsymbol{x}_{u}
        \end{array}
        \right]\\
    \end{aligned}
    \right.
\label{eq_xu}
\end{equation}
where $\boldsymbol{T}$ is defined as the rotational transformation matrix. Each row or column of $\boldsymbol{T}$ contains only one element equal to one and the others equal to zero, and $T_{ij}=1$ means that the $i$-th and the $j$-th elements are centrally symmetric in the grid. In this paper, the 1-elements are arranged on the subdiagonal of the matrix.

The pattern synthesis problem can be then formulated as a minimax problem for minimizing the PSL of $f_{1}\left(\theta,\phi\right)$ and $f_{2}\left(\theta,\phi\right)$ by determining the coding of elements in the binary vector $\boldsymbol{x}_{u}$, which is given by
\begin{equation}
\begin{array}{l}
    {\mathop{\min}\limits_{\boldsymbol{x}_{u}}}\quad 
    \mathop{\max}\{
    |f_{1}\left(\theta,\phi\right)|,
    |f_{2}\left(\theta,\phi\right)|\},
    \left(\theta,\phi\right)\in\Omega_{sl}\\
    s.t.\quad
    \left \{
    \begin{aligned}
        & {{f_{1}}\left( {\theta ,\phi } \right) = \boldsymbol{w}_{h}^{T}\boldsymbol{X}_{1}    \boldsymbol{g}_{hh}(\theta,\phi)}\\
        & {{f_{2}}\left( {\theta ,\phi } \right) = \boldsymbol{w}_{v}^{T}\boldsymbol{X}_{v}    \boldsymbol{g}_{vv}(\theta,\phi)}\\
        & {\boldsymbol{X}_{1}=\text{diag}(\boldsymbol{x}_{1})}\\
        & {\boldsymbol{X}_{2}=\text{diag}(\boldsymbol{x}_{2})}\\
        & \boldsymbol{x}_{1}=
         \left[
        \begin{array}{*{20}{c}}
            \boldsymbol{x}_{u}^{T}&
            \left(\boldsymbol{1}-\boldsymbol{T}\boldsymbol{x}_{u}\right)^{T}
        \end{array}
        \right]^{T} \\
        & \boldsymbol{x}_{2}=\boldsymbol{1}-\boldsymbol{x}_{1}\\
        & x_{u}\in\{0,1\} \\
    \end{aligned}
    \right.
\end{array}
\label{eq_opt}
\end{equation}
This nonlinear integer programming problem is solved using a stochastic optimization algorithm, specifically the BGA adopted in this work. 

For the $i$-th subarray coding vector $\mathbf{x}_i \in \{0,1\}^n$, a transformation matrix $\mathbf{L}_i \in \{0,1\}^{n_i \times n}$ is constructed, where $n_i = \|\mathbf{x}_{i}\|_{1}$ denotes the number of active elements in this subarray. Let $\mathbf{Q} = \{ q_1, q_2, \ldots, q_{n_i} \}$ be the ordered set of indices where the components of $\mathbf{x}_i$ are equal to 1. The transformation matrix $\mathbf{L}_i$ is then defined element-wise by:
\begin{equation}
(\mathbf{L}_i)_{k,l} = \begin{cases}
1, & \text{if } l = q_k \in \mathbf{Q} \\
0, & \text{otherwise}
\end{cases}
\label{eq_Li}
\end{equation}
for row index $k = 1, 2, \ldots, n_i$ and column index $l = 1, 2, \ldots, n$. This matrix maps from the full array's indices to the indices of the given subarray code vector.

Since the dual-beam synthesis problem is decomposed into two independent single-beam synthesis subproblems, for the $i$-th beam ($i=1,2$), the input data can be obtained by mapping the original AEP data $\mathbf{g}_{AB}$ via the transformation matrix $\mathbf{L}_i$:
\begin{equation}
\tilde{\mathbf{g}}_{AB}^{(i)}(\theta,\phi) = \mathbf{L}_{i} \mathbf{g}_{AB}(\theta,\phi)
\label{eq_gABi}
\end{equation}
Given the constraints including the beam direction, polarization state, and sidelobe level, Algorithm \ref{alg:single} is performed to optimize $\tilde{\mathbf{g}}_{AB}^{(i)}(\theta,\phi)$, yielding the local solutions $\mathbf{x}_h^i$ and $\mathbf{x}_v^i$ corresponding to this subarray.

Upon completion of the optimization for all beams, the local solutions are combined into the final code vectors for the full array via the transpose of the transformation matrices:
\begin{equation}
\left \{
\begin{aligned}
\mathbf{x}_{ih} = \mathbf{L}_i^\text{T} \mathbf{x}_h^i \\
\mathbf{x}_{iv} = \mathbf{L}_i^\text{T} \mathbf{x}_v^i
\end{aligned}
\right .
\label{eq_xihxiv}
\end{equation}
The vectors $\mathbf{x}_{ih}$ and $\mathbf{x}_{iv}$ constitute the final output of the dual-beam pattern synthesis process.

Consistent with the approach in Algorithm \ref{alg:single}, the synthesis procedure given by Algorithm \ref{alg:dual} is also founded on the AEP model. By utilizing full-wave simulation or measured data for array pattern calculation, this method inherently accounts for mutual coupling effects. The algorithm steps are explained as follows.

\begin{algorithm}[H]
\caption{The proposed dual-polarized pattern synthesis method for the PCRPA.}\label{alg:dual}
\begin{algorithmic}[1]
\STATE Input the data of active element patterns $\boldsymbol{g}_{AB}(\theta,\phi)$, obtained from either simulations or experiments. Initialize the beam direction $(\theta_0,\phi_0)$ and the sidelobe region $\Omega_{sl}$. Set the termination condition of the run time or the maximum number of generations.
\STATE Optimize the variable $\boldsymbol{x}_{u}$ using the BGA algorithm according to the model given by Equation (\ref{eq_opt}). Stop the process if it reaches the termination condition and obtain the subarray code vectors $\boldsymbol{x}_1$ and $\boldsymbol{x}_2$.
\STATE Perform Algorithm \ref{alg:single} under $\boldsymbol{x}_1$ and $\boldsymbol{x}_2$ respectively according to Equation (\ref{eq_Li})-(\ref{eq_xihxiv}). Return the synthesis results.
\end{algorithmic}
\label{alg2}
\end{algorithm}

\section{Simulation And Analysis}\label{sec4}

This section presents simulation examples to validate the proposed PCRPA architecture and its pattern synthesis methods. The simulation begins with acquiring the AEP data of all ports from a dual-polarized array model through full-wave analysis using the high-frequency structure simulator (HFSS). A dual-polarized microstrip patch antenna, operating at a center frequency of 10 GHz, was employed as the array element. The square patch has a side length of 8.3 mm and is fed by coaxial probes. The array lies in the yoz-plane, with all elements arranged in a $16\times16$ uniform rectangular lattice with an element spacing of 15 mm (i.e., half-wavelength at 10 GHz). Since this section focuses on validating the proposed architecture and methodology, the electromagnetic simulation was configured for a single frequency of 10 GHz. These data are then used as input to the synthesis model, where the binary coding vectors are optimized to emulate the switching functionalities of the architecture in Fig. \ref{fig_bf}.

A set of three key metrics is employed to evaluate the pattern performance: PSL, XPL, and directivity (Dir). PSL and Dir characterize the concentration of co-polarized radiation energy, while XPL quantifies the polarization purity at the beam direction. These metrics are consequently adopted for clear comparison in following simulations and experiments.

\subsection{Arbitrarily Polarized Pattern}

\begin{figure}[htbp]
\centering
\subfloat[]{\includegraphics[width=4cm]{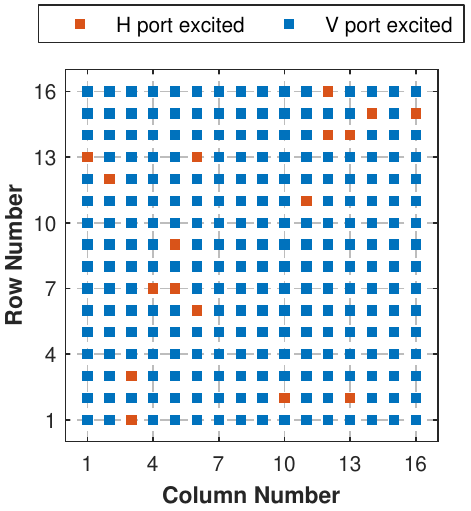}%
\label{fig_exp1_layout}}
\hfil
\subfloat[]{\includegraphics[width=4cm]{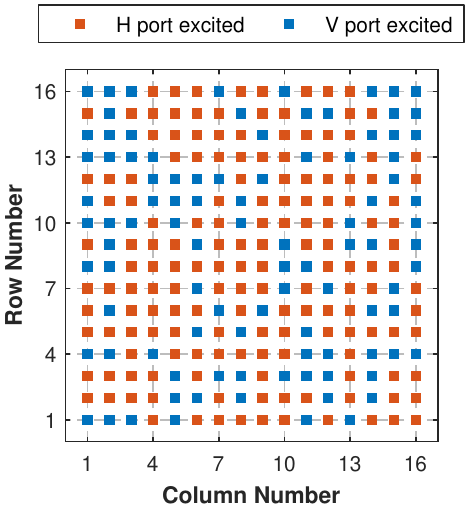}%
\label{fig_exp2_layout}}
\\
\subfloat[]{\includegraphics[width=4.25cm]{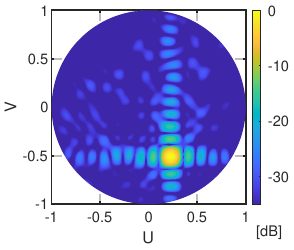}%
\label{fig_exp1_pat_c}}
\hfil
\subfloat[]{\includegraphics[width=4.25cm]{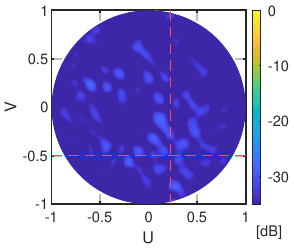}%
\label{fig_exp1_pat_d}}
\\
\subfloat[]{\includegraphics[width=4.25cm]{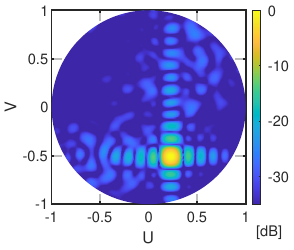}%
\label{fig_exp2_pat_a}}
\hfil
\subfloat[]{\includegraphics[width=4.25cm]{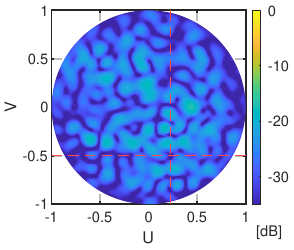}%
\label{fig_exp2_pat_b}}
\caption{Simulation results for arbitrarily polarized pattern synthesis. (a) Array configuration (V-polarization). (b) Array configuration ($45^{\circ}$ linear polarization). (c) CoP pattern of V-polarized beam. (d) XP pattern of V-polarized beam. (e) CoP pattern of $45^{\circ}$ linearly-polarized beam. (f) XP pattern of $45^{\circ}$ linearly-polarized beam.}
\label{fig_exp1_pat}
\end{figure}

This set of simulations evaluates the performance of Algorithm \ref{alg:single} for synthesizing arbitrarily-polarized beams steered to $(120^{\circ},15^{\circ})$. The synthesis results of array configurations and beam patterns for two polarization states, $(90^{\circ},0^{\circ})$ and $(45^{\circ},0^{\circ})$, are presented in Fig. \ref{fig_exp1_pat} as examples.

In Fig. \ref{fig_exp1_layout} and \ref{fig_exp2_layout}, orange and blue colors correspond to the H and V port excitations, respectively. The results show that the PCRPA uses 17 H ports and 239 V ports to synthesize the V-polarized beam, whereas for the $45^{\circ}$ linearly polarized beam, the numbers change to 159 and 97. Both configurations form stable main lobes at the desired direction with controlled PSL and obvious XPL suppression. However, the XP pattern of the $45^{\circ}$ linearly polarized beam exhibits a higher level across the entire space, indicating a sensitivity of the synthesis performance to the polarization state, which will be further discussed later.

\begin{figure}[htbp]
\centering
\subfloat[]{\includegraphics[width=4.25cm]{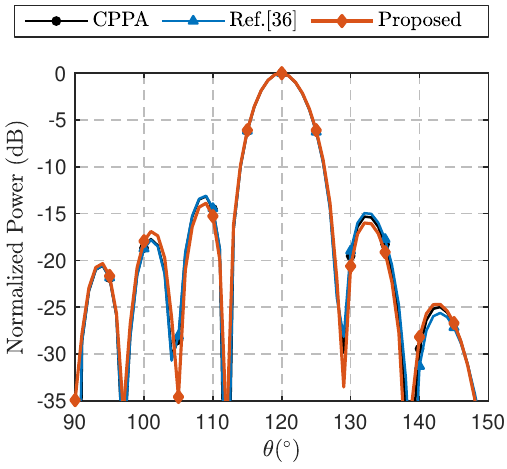}}
\subfloat[]{\includegraphics[width=4.25cm]{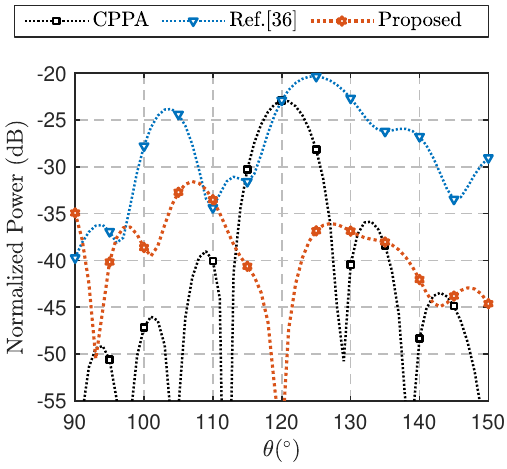}}\\
\subfloat[]{\includegraphics[width=4.25cm]{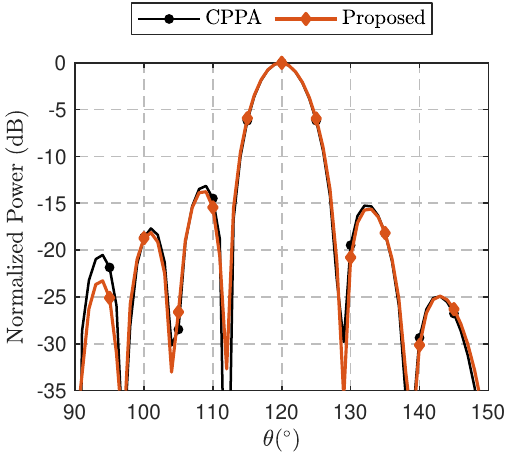}}
\subfloat[]{\includegraphics[width=4.25cm]{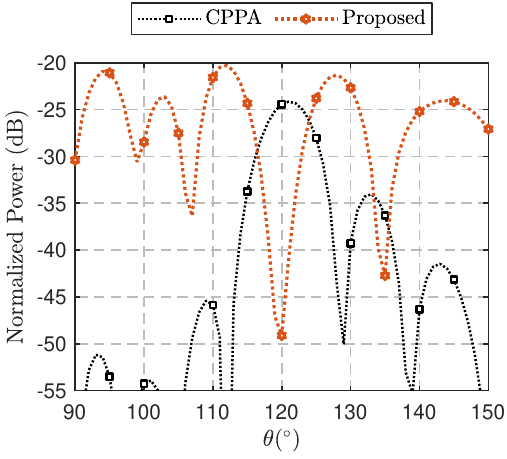}}
\caption{Normalized 2-D power patterns (elevation plane cut) for dual-polarized beams. (a) CoP patterns of V-polarized beams. (b) XP patterns of V-polarized beams. (c) CoP patterns of $45^{\circ}$ linearly-polarized beams. (d) XP patterns of $45^{\circ}$ linearly-polarized beams. }
\label{fig_exp12_comparison}
\end{figure}

Fig. \ref{fig_exp12_comparison} further compares the elevation plane cuts of the PCRPA patterns from Fig. \ref{fig_exp1_pat} with those from the CPPA and \cite{zhou2024}, with the corresponding quantitative metrics summarized in TABLE \ref{tab:single}. The comparative results show that for V-polarized synthesis, both \cite{zhou2024} and the PCRPA maintain a PSL comparable to the CPPA, albeit with directivity loss at approximately 0.2 dB due to elevated integrated sidelobes. In terms of cross-polarization, the architecture in \cite{zhou2024} exhibits an XPL similar to the CPPA, with its XP pattern generally elevated across the entire spatial domain. In contrast, the PCRPA achieves significant suppression of both the XPL and the cross-polarization within the main lobe. For $45^{\circ}$ linearly-polarized synthesis, the PCRPA maintains the same advantages observed in the V-polarized case; however, its level of XP pattern outside the main lobe is further increased. It is noteworthy that the architecture proposed in \cite{zhou2024} cannot form polarization states other than H and V under uniform weighting; therefore, its performance for $45^{\circ}$ linear polarization could not be compared under identical conditions.

\begin{table}[htbp]
    \centering
    \caption{Comparison of key metrics for the arbitrarily polarized pattern synthesis}
    \label{tab:single}
    \begin{tabularx}{\columnwidth}{lcYYY}
        \toprule
        \multicolumn{2}{c}{} & \textbf{CPPA} & \textbf{Ref. \cite{zhou2024}} & \textbf{PCRPA} \\
        \midrule
        \multirow{3}{*}{$(90^{\circ},0^{\circ})$} 
        & XPL (dB) & -22.93 & -22.89 & -59.68 \\
        & PSL (dB) & -12.75 & -12.70 & -13.39 \\
        & Dir (dBi) & 28.37  & 28.14  & 28.15  \\
        \addlinespace
        \multirow{3}{*}{$(45^{\circ},0^{\circ})$} 
        & XPL (dB) & -24.43 & -- & -49.13 \\
        & PSL (dB) & -12.75 & -- & -13.38 \\
        & Dir (dBi) & 28.37  & --  & 28.00  \\
        \addlinespace
        \multicolumn{2}{l}{\textbf{Amplitude Weight}} & Uniform & Uniform & Uniform \\
        \bottomrule
    \end{tabularx}
\end{table}

\subsection{Dual-polarized Patterns}

\begin{figure}[htbp]
\centering
\subfloat[]{\includegraphics[width=5.5cm]{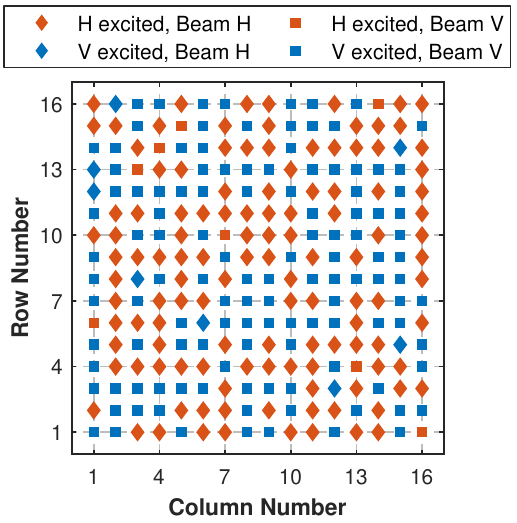}}
\\
\subfloat[]{\includegraphics[width=4.25cm]{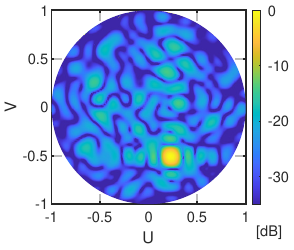}}
\hfil
\subfloat[]{\includegraphics[width=4.25cm]{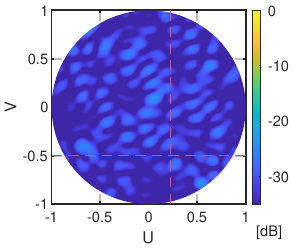}}
\\
\subfloat[]{\includegraphics[width=4.25cm]{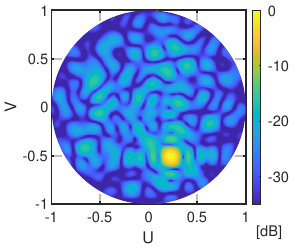}}
\hfil
\subfloat[]{\includegraphics[width=4.25cm]{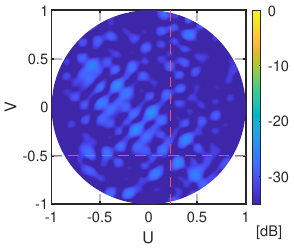}}
\caption{Simulation results for dual-polarized beam synthesis. (a) Array configuration. (b) CoP pattern of H-polarized beam. (c) XP pattern of H-polarized beam. (d) CoP pattern of the V-polarized beam. (e) XP pattern of the V-polarized beam.}
\label{fig_exp3_pat}
\end{figure}

This set of simulations evaluates the performance of the proposed dual-polarized beam synthesis method with beams steered to $(120^{\circ},15^{\circ})$. The sythesis results obtained are presented in Fig. \ref{fig_exp3_pat}. Diamonds and squares denote the subarray excitations for the H- and V-polarized beams, respectively; orange and blue colors correspond to the H and V port excitations of the elements. As shown, each beam is synthesized using half elements of the total array. The H-polarized beam is excited by 120 H ports and 8 V ports, whereas the V-polarized beam is excited by 120 V ports and 8 H ports. Fig \ref{fig_exp3_pat} also presents the CoP and XP patterns for both beams. It can be observed that both CoP patterns form stable main lobes at the desired beam direction, with the PSL below -13 dB, and the XPL suppressed to a low level. Conversely, the integrated sidelobe levels of the CoP patterns are higher than those in the single-beam case, leading to a measurable loss in directivity.

\begin{figure}[htbp]
\centering
\subfloat[]{\includegraphics[width=4.25cm]{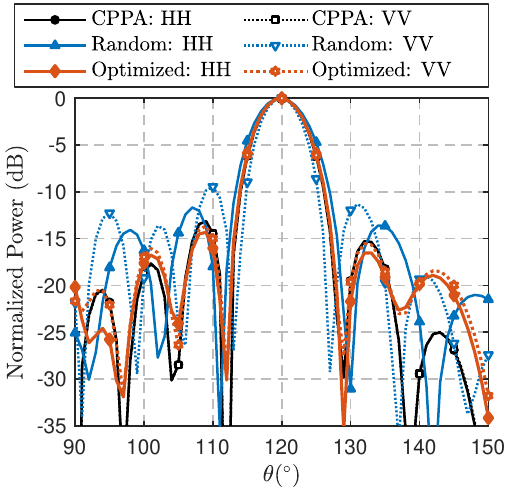}}
\subfloat[]{\includegraphics[width=4.25cm]{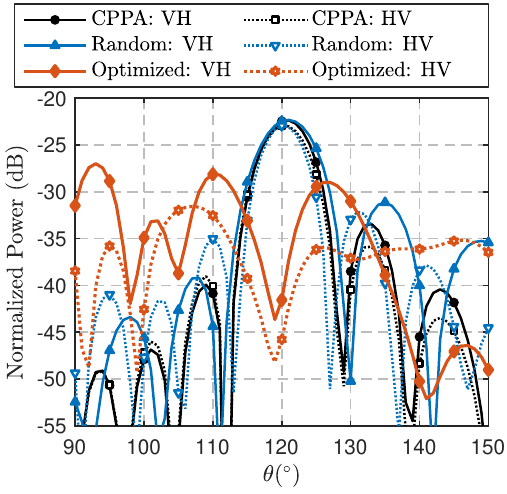}}
\caption{Normalized 2-D power patterns (elevation plane cut) for dual-polarized beams. (a) CoP patterns. (b) XP patterns. }
\label{fig_exp3_comparison}
\end{figure}

Fig. \ref{fig_exp3_comparison} presents the elevation plane cuts of the CoP and XP patterns for the PCRPA dual-beam array from Fig. \ref{fig_exp3_pat}, comparing them with the dual-polarized beams from the CPPA and a randomly coded control group. The corresponding performance metrics are compared in TABLE \ref{tab:dual}. The control group was configured such that its subarray coding vectors, $\boldsymbol{x}_1$ and $\boldsymbol{x}_2$, were randomly generated while maintaining an equal number of excited elements, with $\boldsymbol{x}_1 = \boldsymbol{x}_{1h}$ and $\boldsymbol{x}_2 = \boldsymbol{x}_{2v}$. This configuration excludes the geometric constraints, element position optimization, and polarization port allocation strategies in Algorithm 2. The results demonstrate that the PCRPA optimized by the proposed method maintains comparable PSL while achieving better XPL over the other. The XPL improved from -22.46/-22.93 dB to -41.55/-45.78 dB. Furthermore, compared to the random control group, the proposed algorithm yields well-matched CoP patterns for the H- and V-beams, as evidenced by their nearly identical main lobe widths in Fig. \ref{fig_exp3_comparison}. However, a directivity loss of approximately 2.50 dB is observed in the PCRPA architecture, which is an inherent trade-off due to the use of only a subset of elements for each beam.

\begin{table}[htbp]
    \centering
    \caption{Comparison of key metrics for the dual polarized pattern synthesis}
    \label{tab:dual}
    \begin{tabularx}{\columnwidth}{ccYYYY}
        \toprule
        \multicolumn{2}{c}{\multirow{2}{*}{}} & \multirow{2}{*}{\textbf{CPPA}} & \multicolumn{2}{c}{\textbf{PCRPA}} \\
        \cmidrule(lr){4-5}
        & & & \textbf{random} & \textbf{proposed} \\
        \midrule
        \multirow{2}{*}{XPL (dB)} & H & -22.46 & -22.46 & -41.55 \\
                                & V & -22.93 & -22.94 & -45.78 \\
        \addlinespace
        \multirow{2}{*}{PSL (dB)} & H & -12.73 & -11.68 & -13.31 \\
                                & V & -12.75 & -8.91 & -13.48 \\
        \addlinespace
        \multirow{2}{*}{Dir (dBi)} & H & 28.37  & 25.84  & 25.87 \\
                                 & V & 28.37  & 26.05  & 25.86 \\
        \bottomrule
    \end{tabularx}
\end{table}

\subsection{Parameter Sensitivity Analysis}

\begin{figure*}[htbp]
\centering
\subfloat[]{\includegraphics[width=4.25cm]{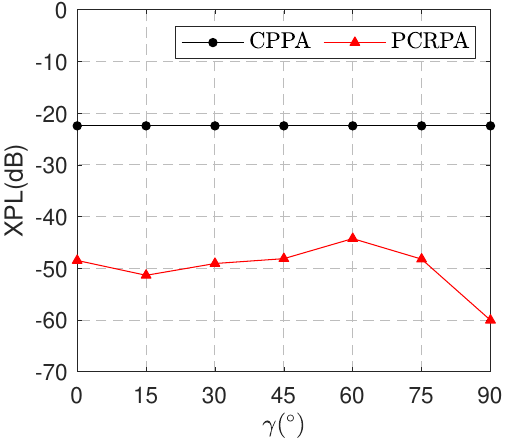}%
\label{fig_exp4_cur_a}}
\hfil
\subfloat[]{\includegraphics[width=4.25cm]{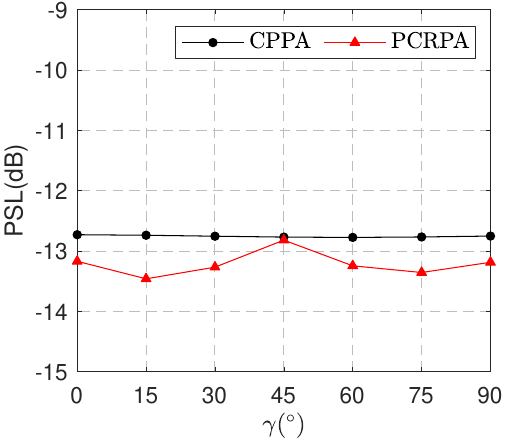}%
\label{fig_exp4_cur_b}}
\hfil
\subfloat[]{\includegraphics[width=4.25cm]{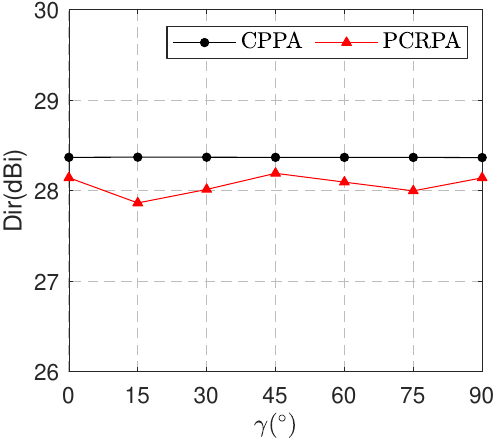}%
\label{fig_exp4_cur_c}}
\hfil
\subfloat[]{\includegraphics[width=4.25cm]{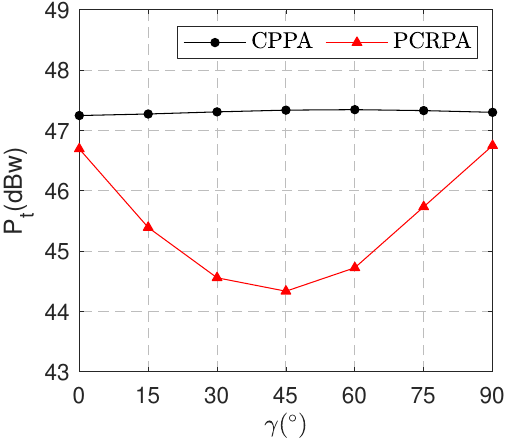}%
\label{fig_exp4_cur_d}}\\
\subfloat[]{\includegraphics[width=4.25cm]{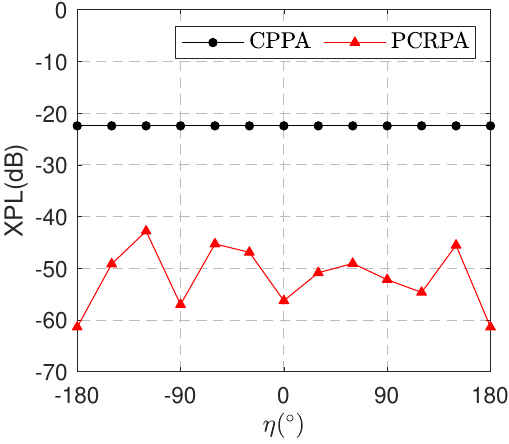}%
\label{fig_exp4_cur_e}}
\hfil
\subfloat[]{\includegraphics[width=4.25cm]{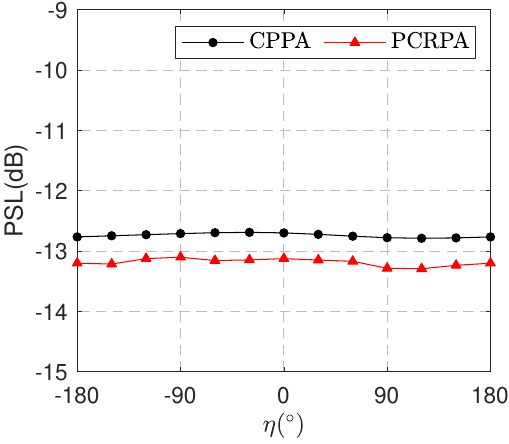}%
\label{fig_exp4_cur_f}}
\hfil
\subfloat[]{\includegraphics[width=4.25cm]{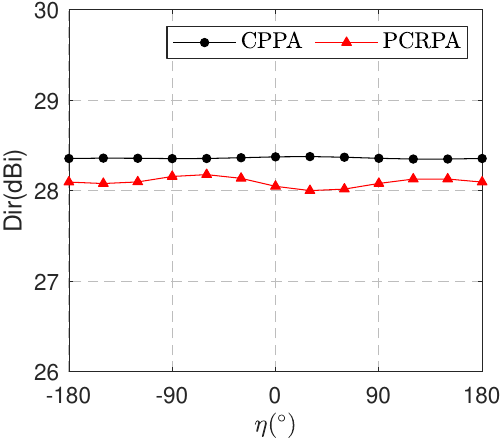}%
\label{fig_exp4_cur_g}}
\hfil
\subfloat[]{\includegraphics[width=4.25cm]{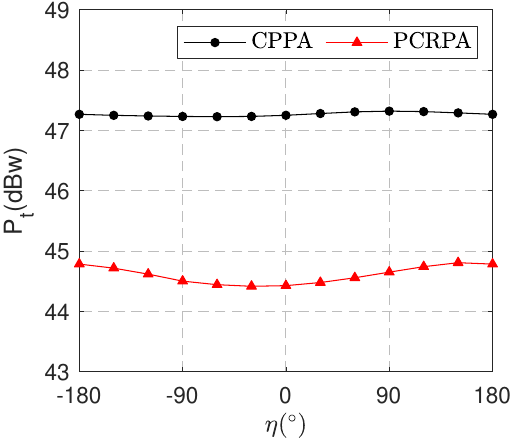}%
\label{fig_exp4_cur_h}}
\caption{Comparison of pattern performance between the conventional and proposed architecture under varying polarization states. (a)-(d) show XPL, PSL, Dir, and $P_t$ as functions of $\gamma$; (e)-(h) present the same metrics as functions of $\eta$.}
\label{fig_exp4_cur}
\end{figure*}

The simulation results have verified the effectiveness of the proposed PCRPA architecture and synthesis methods. This section explores the sensitivity of pattern performance to various input parameters, including the polarization state, beam direction, and array size.

\subsubsection{Polarization State}

This experiment evaluates the impact of polarization parameters on synthesis performance. The polarization states for both the CPPA and PCRPA were synthesized by traversing either $\gamma$ or $\eta$ while keeping the other fixed at $\eta=60^{\circ}$ or $\gamma=30^{\circ}$, respectively, with subsequent pattern metrics collected. To further quantify the sensitivity to polarization parameters, a new variable $P_t$, defined as the radiated power of the CoP pattern at the beam direction, was introduced.

The results given by Fig. \ref{fig_exp4_cur} reveal several key findings. First, the CPPA is largely insensitive to polarization variations, whereas the PCRPA exhibits significant sensitivity. Second, the PCRPA shows low sensitivity to the polarization angle $\eta$; although the XPL curve in Fig. \ref{fig_exp4_cur_e} exhibits fluctuations, all values remain below -40 dB, and the irregular variations can reasonably be attributed to differences in active element patterns under mutual coupling. Crucially, the PCRPA's performance metrics—PSL, Dir, and $P_t$—are highly sensitive to the polarization angle $\gamma$. As shown in Figs. \ref{fig_exp4_cur_b}-\ref{fig_exp4_cur_d}, when $\gamma$ approaches $0^{\circ}$, $45^{\circ}$, or $90^{\circ}$, the patterns display higher PSL and Dir. However, a distinct $P_t$ loss of approximately 3 dB occurs near $\gamma=45^{\circ}$, while this loss diminishes to nearly zero as $\gamma$ approaches $0^{\circ}$ or $90^{\circ}$.

This behavior originates from the synthesis mechanism of the PCRPA: the ratio of H-to-V ports is determined by $\gamma$ given by Equation (\ref{eq_num}). When this ratio approaches 1, or when ports are entirely H or V, the PCRPA's CoP pattern most closely resembles that of the CPPA, leading to similar PSL and Dir. However, the vector superposition of orthogonal radiation fields from a nearly equal number of ports results in energy cancellation during polarization synthesis, which causes the observed $P_t$ degradation. This warrants careful consideration in low SNR environments where power efficiency is critical for detection accuracy.

\subsubsection{Array Size}

\begin{figure*}[htbp]
\centering
\subfloat[]{\includegraphics[width=5.5cm]{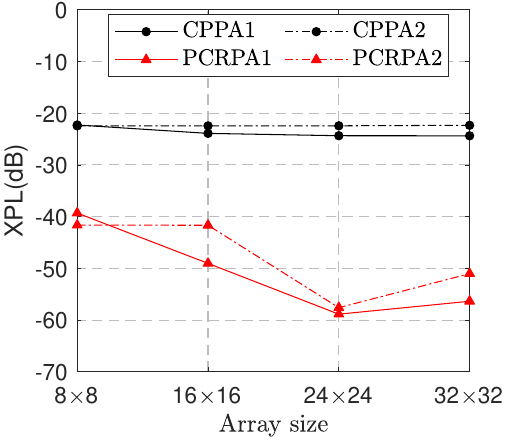}%
\label{fig_exp6_cur_a}}
\hfil
\subfloat[]{\includegraphics[width=5.5cm]{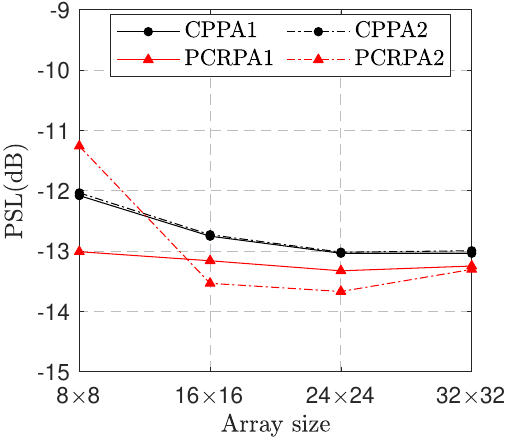}%
\label{fig_exp6_cur_b}}
\hfil
\subfloat[]{\includegraphics[width=5.5cm]{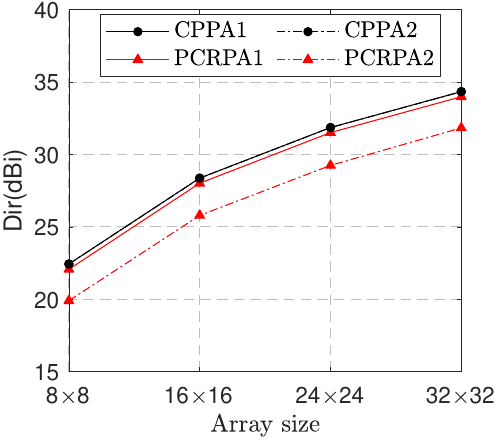}%
\label{fig_exp6_cur_c}}
\caption{Comparison of pattern performance between the conventional and proposed arrays across different array sizes. The legend distinguishes the beam types: '1' for single-polarized and '2' for dual-polarized beams. (a) XPL. (b) PSL. (c) Directivity.}
\label{fig_exp6_cur}
\end{figure*}

\begin{figure*}[htbp]
\centering
\subfloat[]{\includegraphics[width=5.5cm]{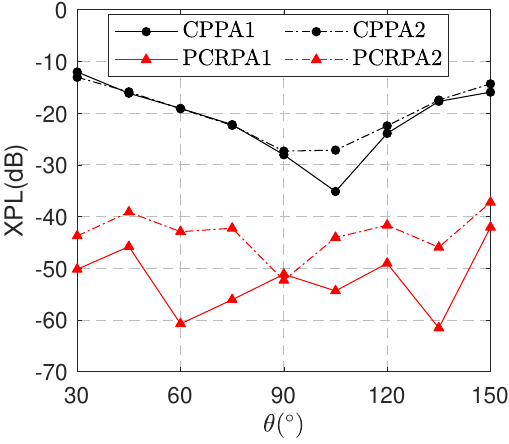}%
\label{fig_exp5_cur_a}}
\hfil
\subfloat[]{\includegraphics[width=5.5cm]{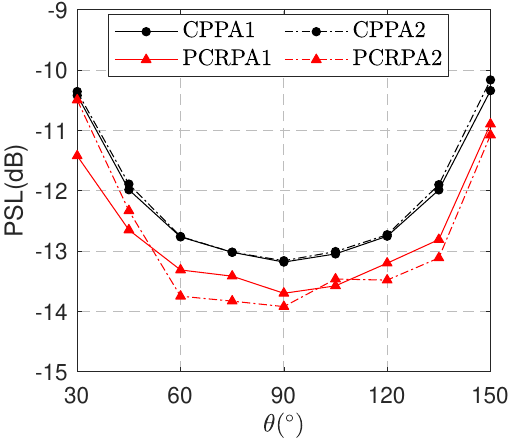}%
\label{fig_exp5_cur_b}}
\hfil
\subfloat[]{\includegraphics[width=5.5cm]{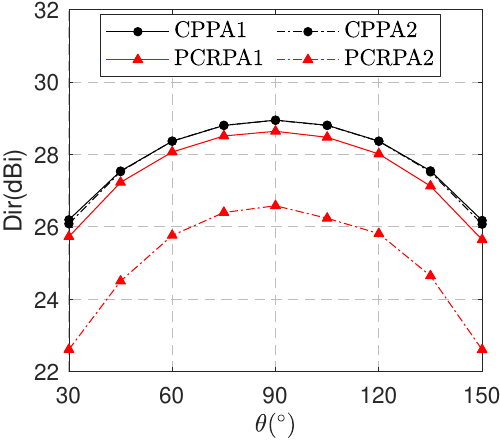}%
\label{fig_exp5_cur_c}}
\caption{Comparison of pattern performance between the conventional and proposed arrays across different beam directions. The legend distinguishes the beam types: '1' for single-polarized and '2' for dual-polarized beams. (a) XPL. (b) PSL. (c) Directivity.}
\label{fig_exp5_cur}
\end{figure*}

This experiment investigates the impact of array size on synthesis performance. With the beam direction and polarization fixed, arrays of sizes $8\times8$, $16\times16$, $24\times24$, and $32\times32$ were evaluated. The AEPs of these arrays served as the input for both CPPA and PCRPA in arbitrary-polarized and dual-polarized pattern synthesis, with subsequent pattern metrics calculated.

The results given by Fig. \ref{fig_exp6_cur} reveals several key trends. First, as shown in Fig. \ref{fig_exp6_cur_a}, the XPL of the PCRPA decreases rapidly as the array size increases from $8\times8$ to $24\times24$, plateauing after falling below -50 dB. Second, Fig. \ref{fig_exp6_cur_b} indicates that the PSL for dual-beam synthesis in the PCRPA improves significantly when scaling from $8\times8$ to $16\times16$, then stabilizes, whereas its single-beam PSL declines gradually throughout. Furthermore, the directivity of both CPPA and PCRPA increases with array size, following similar growth trends.

\begin{figure}[htbp]
\centering
\includegraphics[width=6cm]{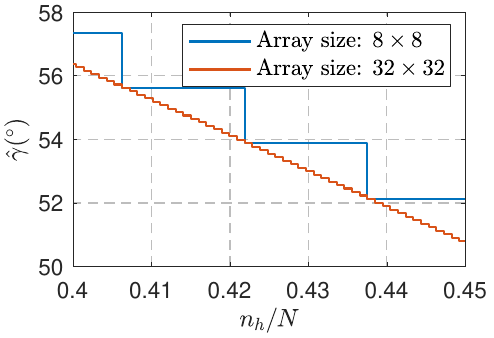}
\caption{The quantization curves of $\hat{\gamma}$ concerning different array size.}
\label{fig_exp6_quant}
\end{figure}

These findings indicate that the XPL and dual-beam PSL of the PCRPA are sensitive to the array size. The sensitivity of the dual-beam PSL shares a similar principle with thinned array synthesis, primarily governed by the dimensionality of the optimization problem. The sensitivity of the XPL, however, is inherent to its suppression mechanism of the PCRPA. The polarization angle $\hat{\gamma}$ in the PCRPA is a discrete, quantized value when mutual coupling is neglected, which can be expressed by:
\begin{equation}
    \hat{\gamma}=\arctan (\frac{n_v}{n_h})
    \label{eq_gamma}
\end{equation}
A comparison of the $\hat{\gamma}$ quantization curves for the $8\times8$ and $32\times32$ arrays in Fig. \ref{fig_exp6_quant} clearly shows that a larger array results in a finer quantization step. This enhances the accuracy of approximating the desired polarization state, thereby improving polarization purity and yielding a lower XPL.

\subsubsection{Beam Direction}

This experiment evaluates the impact of beam direction. With the azimuth angle fixed at $\phi = 30^\circ$, the elevation angle $\theta$ was swept from $30^\circ$ to $150^\circ$ to define the beam direction for both CPPA and PCRPA in arbitrary-polarized and dual-polarized pattern synthesis. The corresponding pattern metrics were recorded.

The results given by Fig. \ref{fig_exp5_cur} reveal several key observations. First, the CPPA's XPL, PSL, and directivity all gradually degrade with beam scanning. Second, as shown in Figs. \ref{fig_exp5_cur_b} and \ref{fig_exp5_cur_c}, the PSL and directivity of the PCRPA follow the same trend as the CPPA. Although the proposed algorithm maintains PSL at a lower level than the CPPA, a consistent directivity loss is observed—approximately 0.3 dB for single-beam and 2.5 dB for dual-beam synthesis. Finally, Fig. \ref{fig_exp5_cur_a} demonstrates that the PCRPA achieves significant XPL suppression, with the improvement becoming more pronounced at larger scanning angles. The single-beam synthesis yields better XPL suppression compared to the dual-beam case; however, the XPL values themselves fluctuate without a clear correlation to the scanning angle.

The similarity in PSL and directivity trends between the PCRPA and CPPA is attributed to their shared dependence on the element pattern. The XPL suppression in single-beam synthesis occurs because all array DoFs are allocated to synthesizing a single polarization state, resulting in finer equivalent polarization port quantization. This principle aligns with the array size sensitivity analysis discussed previously.

\section{Experiment and Discussion}\label{sec5}
\subsection{Experimental Results}

\begin{figure*}[htbp]
\centering
\subfloat[]{\includegraphics[width=7cm]{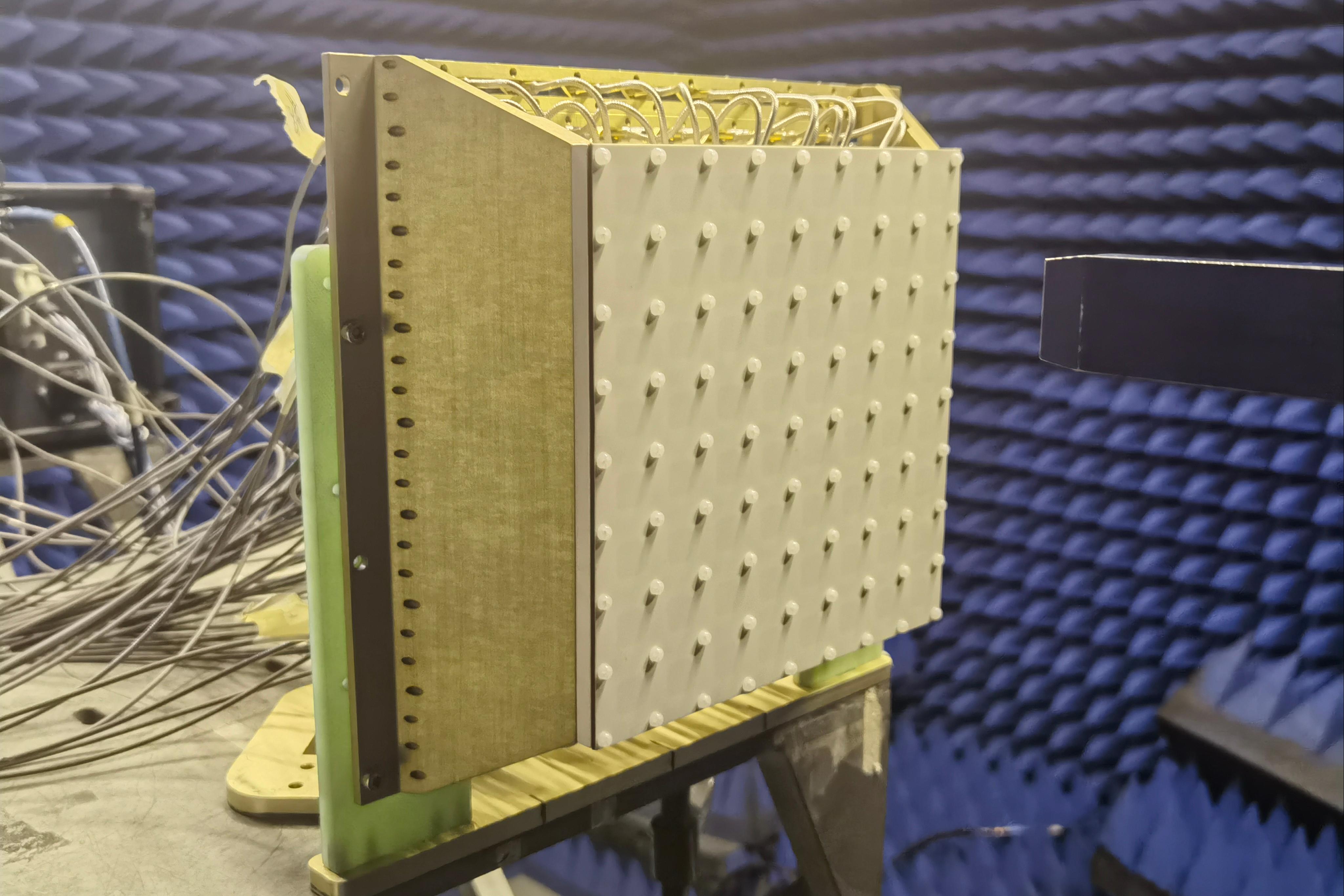}%
\label{fig_antenna_a}}
\hfil
\subfloat[]{\includegraphics[width=7cm]{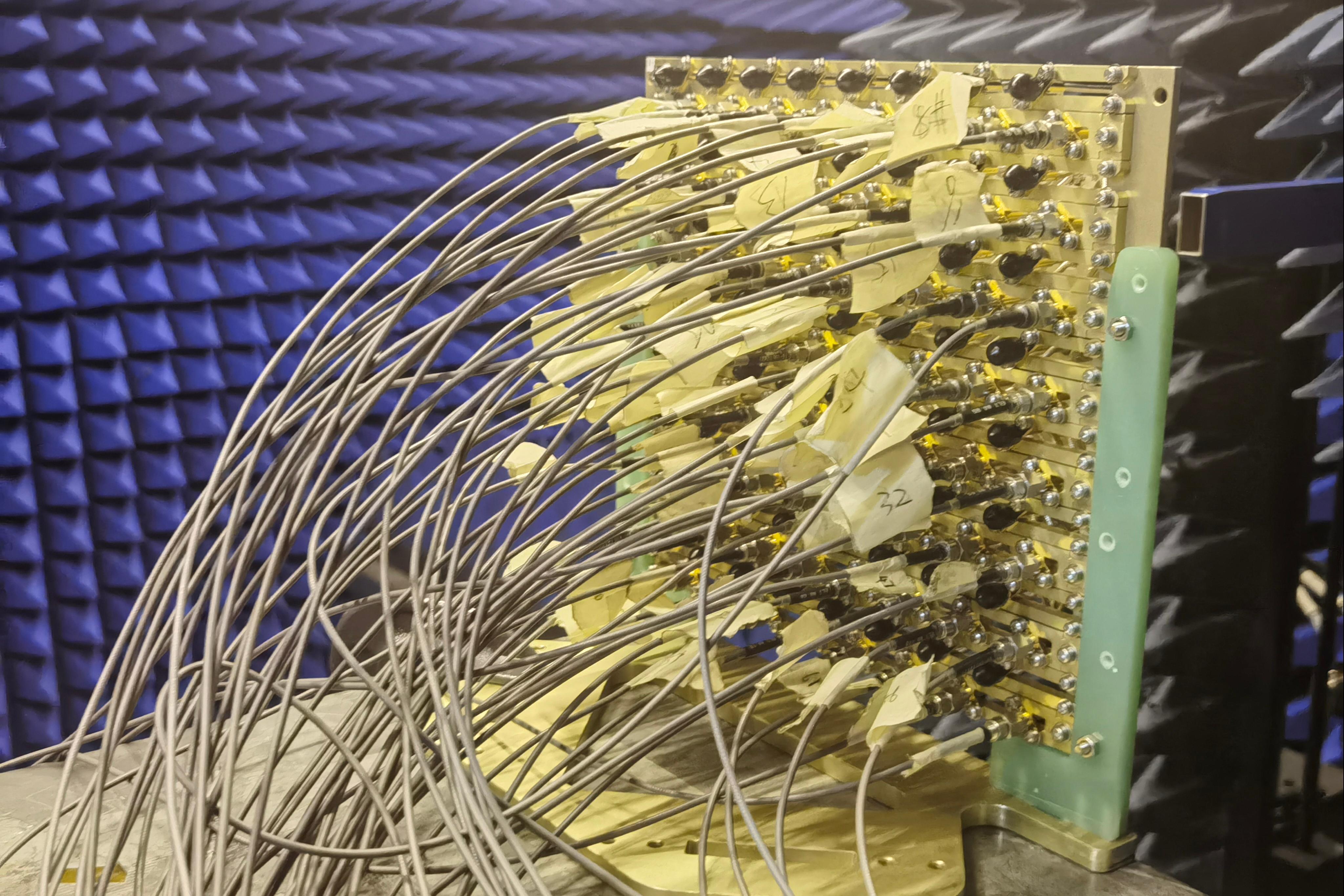}%
\label{fig_antenna_b}}
\\
\caption{A standard commercial X-band $8\times 8$-unit dual-polarized microstrip patch array antenna test in an anechoic chamber. (a) Front view. (b) Back view.}
\label{fig_antenna}
\end{figure*}

To verify the effectiveness of the proposed architecture of the PCRPA and synthesis methods, an experiment is carried out using a standard commercial dual-polarized microstrip patch antenna, shown in Fig. \ref{fig_antenna}. The antenna operates at a frequency of 9.3 GHz, with a bandwidth of 200 MHz. The array size of $8\times8$, and the element spacing is 16.1 mm. A planar near-field measurement was conducted for the antenna. During the test, the polarized ports were connected via a switch matrix to a vector network analyzer through independent channels, as shown in Fig. \ref{fig_antenna_b}. For all these ports, the 3-D AEPs were acquired via probe scanning and near-field to far-field transformation as measured results, which were then used as the input in the simulation platform. By controlling the 0-1 coding and phase weighting corresponding to each data set, the physical functions of the RF switches and phase shifters given in Fig. \ref{fig_bf} were emulated, thereby validating the principle of the proposed architecture and methodology.

\begin{figure}[htbp]
\centering
\includegraphics[width=7.2cm]{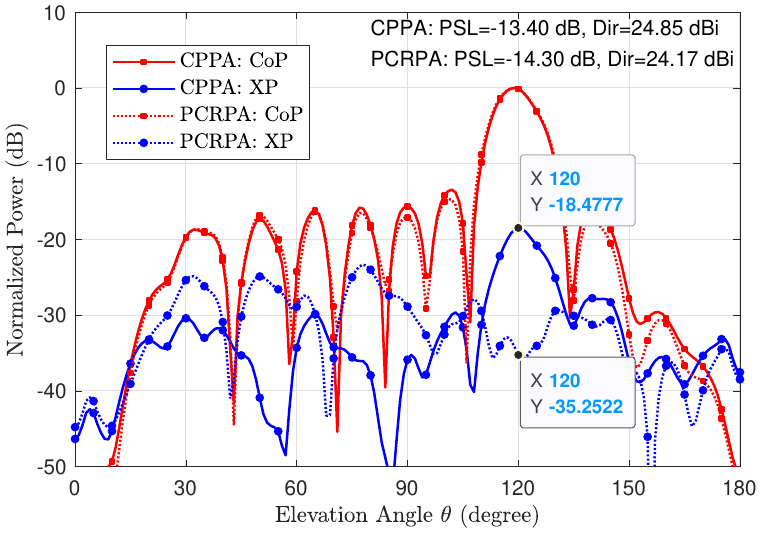}
\caption{Measured CoP and XP patterns (elevation cut) under the conventional and proposed architecture for single polarized pattern synthesis.}
\label{fig_exp7_pat1}
\end{figure}

Fig. \ref{fig_exp7_pat1} compares the CoP and XP patterns of the conventional and proposed architectures for single-polarized pattern synthesis, under the same beam direction $(120^{\circ},15^{\circ})$ and V-polarization setting as used in simulations. The measured results align well with the simulation trends: the PCRPA achieves a significant XPL suppression from -18.48 dB (CPPA) to -35.26 dB, an improvement of approximately 16.78 dB. While this value is slightly higher than the simulated -40 dB level in Fig. \ref{fig_exp6_cur_b}, it still supports the conclusion that the proposed architecture effectively suppresses XPL. Furthermore, the PSL remains nearly identical between the two architectures (-13.40 dB for CPPA vs. -14.30 dB for PCRPA), and a limited directivity loss of 0.68 dB is observed in the PCRPA. These experimental findings—XPL suppression, maintained PSL, and a minor directivity trade-off—collectively validate the key performance trends predicted by the simulations.

\begin{figure}[htbp]
\centering
\subfloat[]{\includegraphics[width=7.2cm]{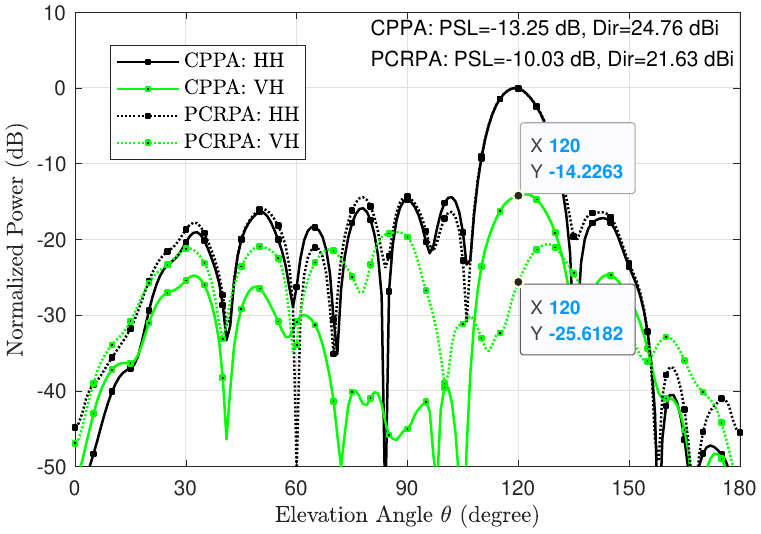}%
\label{fig_exp7_pat2_a}}
\\
\subfloat[]{\includegraphics[width=7.2cm]{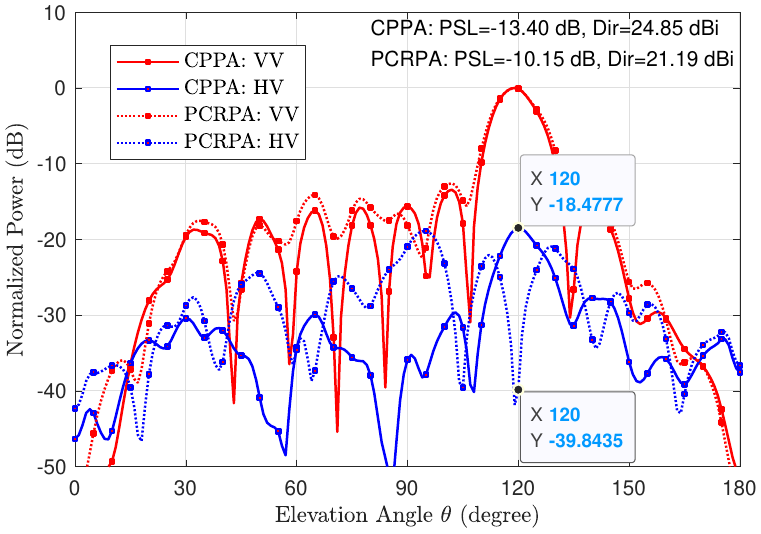}%
\label{fig_exp7_pat2_b}}
\\
\caption{Measured patterns (elevation cut) under different architectures for dual polarized pattern synthesis. (a) H-polarized beam. (b) V-polarized beam. }
\label{fig_exp7_pat2}
\end{figure}

Fig. \ref{fig_exp7_pat2} presents the measured 2-D power patterns for dual-polarized synthesis under the conventional and proposed architectures, also with the beam direction set to $(120^{\circ},15^{\circ})$. The experimental outcomes confirm the expected trade-offs: the PCRPA effectively suppresses XPL, reducing it from -14.23/-18.48 dB to -25.62/-39.84 dB through element polarization port reconfiguration. However, as also indicated in simulations, this improvement is accompanied by a PSL increase of 3.25 dB and a more noticeable directivity loss of 3.66 dB, attributable to the reduction in active channels. It is observed that the H-polarized beam exhibits less XPL suppression than its V-polarized counterpart, a phenomenon we attribute to the heightened sensitivity of mutual coupling, fabrication and environmental variations in smaller arrays. It is anticipated that this performance would improve with larger array scales, as stated in the sensitivity analysis, where such practical factors exert a relatively smaller influence. Overall, these results align with simulated predictions while underscoring the impact of real-world constraints on achievable performance.

\subsection{Discussion}

Analysis of both simulation and experimental results shows that the array size is a crucial factor for the performance of the proposed PCRPA architecture. Employing PCRPAs in larger arrays is recommended. As the array size increases, the proposed PCRPAs exhibit lower sidelobe levels and better control over cross-polarization levels of the patterns. This can be attributed to several reasons:

\begin{enumerate}
 \item The PCRPA architecture is subject to quantization errors, which decrease as the array size increases. This kind of error will limit the effects of XPL mitigation as the desired polarization state cannot be precisely synthesized. Larger arrays allow for finer resolution, thereby reducing the impact of quantization.
 \item The pattern synthesis for PCRPAs is affected by various practical factors such as mutual coupling. The increase in the array size tends to mitigate these effects, leading to more precise and desirable pattern synthesis results.
 \item For the dual polarized pattern synthesis, larger arrays can offer more DoFs in the thinned array synthesis problem, allowing for higher limits on the PSL of the CoP patterns.
\end{enumerate}

Furthermore, this architecture's main advantage is its low cost compared to the conventional design. When the array size increases, the cost contradiction becomes more apparent. This also determines that the proposed PCRPA architecture is more suitable for addressing the cost-performance ratio issues in large-scale polarimetric phased array systems.

\section{Conclusion}\label{sec6}
An architecture called PCRPA and its efficient pattern synthesis methods are proposed. Compared to the conventional architecture, the number of T/R channels in PCRPA is reduced by half using polarization coding reconfigurable elements. By simultaneously controlling the code and weight of each element, synthesis methods obtain arbitrarily or dual-polarized beam patterns with required beam directions, PSL, and XPL in an acceptable running time. This provides a low-cost scheme for PPA radars or other applications to save costs while ensuring beamforming abilities.

Despite the advantages, simulation results show that the proposed technique exhibits sensitivity to array size in terms of PSL and XPL in the pattern synthesis. Especially for XPL, the mitigation effect is influenced by multiple factors, such as quantization errors and mutual coupling. Moreover, due to the reduction in the number of T/R channels, there is an inevitable loss in beam power and directivity.

Looking forward, future research will address the identified limitations in PSL and XPL performance. Additionally, extending the proposed architecture to conformal arrays presents an important research direction. In such geometries, position-dependent polarization variations and complex mutual coupling effects due to surface curvature introduce significant challenges. Future work will explore modeling strategies and algorithmic adaptations to handle these problems.

\bibliographystyle{IEEEtran}
\bibliography{main}

@ARTICLE{zhu2024principles,
  author={Zhu, Di and Xu, Zhiming and Li, Nanjun and Wang, Fulai and Pang, Chen and Li, Yongzhen and Wang, Xuesong},
  journal={IEEE Transactions on Geoscience and Remote Sensing}, 
  title={Principles and Methods of Radar Super-Resolution Based on Instantaneous Polarization Response}, 
  year={2023},
  volume={61},
  number={},
  pages={1-16},
  doi={10.1109/TGRS.2023.3323411}}

@ARTICLE{wu2024a,
  author={Wu, Guoqing and Wang, Shengbin Luo and Wang, Ping and Li, Yongzhen},
  journal={IEEE Transactions on Aerospace and Electronic Systems}, 
  title={A Polarization-Doppler Joint Feature-Based Detection Method for Small Targets in Sea Clutter}, 
  year={2024},
  volume={60},
  number={6},
  pages={8791-8804},
  doi={10.1109/TAES.2024.3433328}}

@ARTICLE{compton1981on,
  author={Compton, R.},
  journal={IEEE Transactions on Antennas and Propagation}, 
  title={On the performance of a polarization sensitive adaptive array}, 
  year={1981},
  volume={29},
  number={5},
  pages={718-725},
  doi={10.1109/TAP.1981.1142651}}

@ARTICLE{lu2021adaptive,
  author={Lu, Yawei and Ma, Jiazhi and Zhou, Jian and Shi, Longfei},
  journal={IEEE Antennas and Wireless Propagation Letters}, 
  title={Adaptive Polarization-Constrained Monopulse Approach for Dual-Polarization Array}, 
  year={2021},
  volume={20},
  number={3},
  pages={289-292},
  doi={10.1109/LAWP.2020.3047684}}

@ARTICLE{zhou2024,
  author={Zhou, Jian and Wang, Zhanling and Wang, Yiqing and Zhu, Di and Pang, Chen and Li, Yongzhen},
  journal={IEEE Transactions on Antennas and Propagation}, 
  title={Polarization-Reconfigurable Phased Array Architecture With Optimally Polarized Elements}, 
  year={2025},
  volume={73},
  number={1},
  pages={201-215},
  doi={10.1109/TAP.2024.3465851}}

@ARTICLE{salazar2024,
  author={Salazar, Cesar M. and Schvartzman, David and Cheong, Boonleng and Palmer, Robert D.},
  journal={IEEE Transactions on Geoscience and Remote Sensing}, 
  title={A Novel Cross-Polar Canceller Technique for Improved Polarimetric Performance of Fully Digital Phased Array Radar}, 
  year={2024},
  volume={62},
  number={},
  pages={1-17},
  doi={10.1109/TGRS.2024.3435520}}

@ARTICLE{wang2022,
  author={Wang, Yan and Xu, Feng and Jin, Ya-Qiu and Du, Zhengwei},
  journal={IEEE Transactions on Antennas and Propagation}, 
  title={Low-Cost Reconfigurable 1 bit Millimeter-Wave Array Antenna for Mobile Terminals}, 
  year={2022},
  volume={70},
  number={6},
  pages={4507-4517},
  doi={10.1109/TAP.2022.3140508}}

@article{chen2022,
title = {A Polarization Programmable Antenna Array},
journal = {Engineering},
volume = {16},
pages = {100-114},
year = {2022},
issn = {2095-8099},
doi = {https://doi.org/10.1016/j.eng.2022.03.015}                ,
url = {https://www.sciencedirect.com/science/article/pii/S2095809922002843},
author = {Dingzhao Chen and Yanhui Liu and Ming Li and Pan Guo and Zhuo Zeng and Jun Hu and Y. Jay Guo}
}

@ARTICLE{wang2024low-cost,
  author={Wang, Yiqing and Zhou, Jian and Wang, Zhanling and Pang, Chen and Li, Yongzhen and Wang, Xuesong},
  journal={IEEE Antennas and Wireless Propagation Letters}, 
  title={Low-Cost Architecture Using Interleaved Thinning for Polarimetric Phased Array}, 
  year={2024},
  volume={23},
  number={3},
  pages={980-984},
  doi={10.1109/LAWP.2023.3340860  }}

@article{fulton2016digital,
  title={Digital phased arrays: Challenges and opportunities},
  author={Fulton, Caleb and Yeary, Mark and Thompson, Daniel and Lake, John and Mitchell, Adam},
  journal={Proceedings of the IEEE},
  volume={104},
  number={3},
  pages={487--503},
  year={2016},
  publisher={IEEE}
}

@ARTICLE{palmer2023horus,
  author={Palmer, Robert D. and Yeary, Mark B. and Schvartzman, David and Salazar-Cerreno, Jorge L. and Fulton, Caleb and McCord, Matthew and Cheong, Boonleng and Bodine, David and Kirstetter, Pierre and Sigmarsson, Hjalti H. and Yu, Tian-You and Zrnić, Dušan and Kelley, Redmond and Meier, John and Herndon, Matthew},
  journal={IEEE Transactions on Radar Systems}, 
  title={Horus—A Fully Digital Polarimetric Phased Array Radar for Next-Generation Weather Observations}, 
  year={2023},
  volume={1},
  number={},
  pages={96-117},
  doi={10.1109/TRS.2023.3280033}}

@ARTICLE{wan2023code,
  author={Wan, Xiang and Li, Bai Yang and Wang, Wen Hao and Huang, Zi Ai and Wang, Jia Wei and Wang, Xu Jie and Cheng, Qiang and Cui, Tie Jun},
  journal={IEEE Transactions on Antennas and Propagation}, 
  title={Code-Dependent Beam Shaping and Polarization Control by Binary Transmitarray}, 
  year={2023},
  volume={71},
  number={6},
  pages={4834-4842},
  doi={10.1109/TAP.2023.3265460}}

@ARTICLE{barbetty2012,
  author={Sánchez-Barbetty, Mauricio and Jackson, Robert W. and Frasier, Stephen},
  journal={IEEE Transactions on Geoscience and Remote Sensing}, 
  title={Interleaved Sparse Arrays for Polarization Control of Electronically Steered Phased Arrays for Meteorological Applications}, 
  year={2012},
  volume={50},
  number={4},
  pages={1283-1290},
  doi={10.1109/TGRS.2011.2167016}}

@ARTICLE{salazar2024a,
  author={Salazar, Cesar M. and Schvartzman, David and Cheong, Boonleng and Palmer, Robert D.},
  journal={IEEE Transactions on Geoscience and Remote Sensing}, 
  title={A Novel Cross-Polar Canceller Technique for Improved Polarimetric Performance of Fully Digital Phased Array Radar}, 
  year={2024},
  volume={62},
  number={},
  pages={1-17},
  doi={10.1109/TGRS.2024.3435520}}

@INPROCEEDINGS{zhang2022cylindrical,
  author={Zhang, Guifu},
  booktitle={2022 IEEE Radar Conference (RadarConf22)}, 
  title={Cylindrical Polarimetric Phased Array Radar for Weather Observations: An Review}, 
  year={2022},
  volume={},
  number={},
  pages={01-05},
  doi={10.1109/RadarConf2248738.2022.9764328}}

@ARTICLE{pan2023joint,
  author={Pan, Bunian and Dong, Luxin and Yu, Xianxiang and Yao, Xue and Sha, Minghui and Cui, Guolong},
  journal={IEEE Sensors Journal}, 
  title={Joint Polarization-Space-Time Processing for Mainlobe Jamming via CP Decomposition}, 
  year={2023},
  volume={23},
  number={13},
  pages={14781-14794},
  doi={10.1109/JSEN.2023.3279069}}

@ARTICLE{fulton2017cylindrical,
  author={Fulton, Caleb and Salazar, Jorge L. and Zhang, Yan and Zhang, Guifu and Kelly, Redmond and Meier, John and McCord, Matt and Schmidt, Damon and Byrd, Andrew D. and Bhowmik, Lal Mohan and Karimkashi, Shaya and Zrnic, Dusan S. and Doviak, Richard J. and Zahrai, Allen and Yeary, Mark and Palmer, Robert D.},
  journal={IEEE Transactions on Geoscience and Remote Sensing}, 
  title={Cylindrical polarimetric phased array radar: Beamforming and calibration for weather applications}, 
  year={2017},
  volume={55},
  number={5},
  pages={2827-2841},
  doi={10.1109/TGRS.2017.2655023}}

@ARTICLE{herd2016the,
  author={Herd, Jeffrey S. and Conway, M. David},
  journal={Proceedings of the IEEE}, 
  title={The Evolution to Modern Phased Array Architectures}, 
  year={2016},
  volume={104},
  number={3},
  pages={519-529},
  doi={10.1109/JPROC.2015.2494879 }}

@ARTICLE{hurtado2008polar,
  author={Hurtado, Martin and Nehorai, Arye},
  journal={IEEE Transactions on Signal Processing}, 
  title={Polarimetric Detection of Targets in Heavy Inhomogeneous Clutter}, 
  year={2008},
  volume={56},
  number={4},
  pages={1349-1361},
  doi={10.1109/TSP.2007.909046 }}

@article{cheng2015antenna, 
  author = {Cheng, Xu and Li, Yong-zhen and Wang, Xue-song}, 
  title = {Antenna Polarization Optimization for Target Detection in Non-Gaussian Clutter}, 
  journal = {International Journal of Antennas and Propagation}, 
  volume = {2015}, 
  number = {1}, 
  pages = {640730},
  year = {2015},
}

@ARTICLE{rocca2016uncon,
  author={Rocca, Paolo and Oliveri, Giacomo and Mailloux, Robert J. and Massa, Andrea},
  journal={Proceedings of the IEEE}, 
  title={Unconventional Phased Array Architectures and Design Methodologies—A Review}, 
  year={2016},
  volume={104},
  number={3},
  pages={544-560},
  doi={10.1109/JPROC.2015.2512389 }}

@INPROCEEDINGS{salazar2015transmit,
  author={Salazar, Jorge L. and Medina, Rafel H. and Loew, Eric},
  booktitle={2015 IEEE MTT-S International Microwave Symposium}, 
  title={Transmit/receive (T/R) modules architectures for dual-polarized weather phased array radars}, 
  year={2015},
  volume={},
  number={},
  pages={1-4},
  doi={10.1109/MWSYM.2015.7167077}}

@INPROCEEDINGS{salazar2015tr,
  author={Salazar, Jorge L. and Medina, Rafel H. and Loew, Eric},
  booktitle={2015 IEEE Radar Conference (RadarCon)}, 
  title={T/R modules for active phased array radars}, 
  year={2015},
  volume={},
  number={},
  pages={1125-1133},
  doi={10.1109/RADAR.2015.7131163}}

@article{mailloux2009irreg,
  author = {Mailloux, R. J. and Santarelli, S. G. and Roberts, T. M. and Luu, D.}, 
  title = {Irregular Polyomino-Shaped Subarrays for Space-Based Active Arrays}, 
  journal = {International Journal of Antennas and Propagation}, 
  volume = {2009}, 
  number = {1}, 
  pages = {956524}, 
  doi = {https://doi.org/10.1155/2009/956524            }     , 
  year = {2009}
}

@ARTICLE{zhu2024a,
  author={Zhu, Di and Wang, Fulai and Zhou, Jian and Li, Nanjun and Wang, Taoran and Pang, Chen and Li, Yongzhen},
  journal={IEEE Transactions on Aerospace and Electronic Systems}, 
  title={A Range Super-resolution Scheme Based on Polarimetric Partially Coherent Radar}, 
  year={2024},
  volume={},
  number={},
  pages={1-19},
  doi={10.1109/TAES.2024.3457929 }}

@ARTICLE{liu2024robust,
  author={Liu, Yibin and Wang, Shengbin Luo and Wu, Guoqing and Wang, Ping and Li, Yongzhen},
  journal={IEEE Transactions on Signal Processing}, 
  title={Robust and Unambiguous Four-Channel Monopulse Two-Target Resolution: A Polarimetric Closed-Form Approach}, 
  year={2024},
  volume={72},
  number={},
  pages={4222-4236},
  doi={10.1109/TSP.2024.3452239   }}

@ARTICLE{mirmozafari2019dual,
  author={Mirmozafari, Mirhamed and Zhang, Guifu and Fulton, Caleb and Doviak, Richard J.},
  journal={IEEE Antennas and Propagation Magazine}, 
  title={Dual-Polarization Antennas With High Isolation and Polarization Purity: A Review and Comparison of Cross-Coupling Mechanisms}, 
  year={2019},
  volume={61},
  number={1},
  pages={50-63},
  doi={10.1109/MAP.2018.2883032   }}

@ARTICLE{ta2015crossed,
  author={Ta, Son Xuat and Park, Ikmo and Ziolkowski, Richard W.},
  journal={IEEE Antennas and Propagation Magazine}, 
  title={Crossed Dipole Antennas: A review}, 
  year={2015},
  volume={57},
  number={5},
  pages={107-122},
  doi={10.1109/MAP.2015.2470680   }}

@ARTICLE{an2024ground,
  author={An, Mengyun and Yin, Jiapeng and Huang, Jiankai and Tan, Xue and Li, Yongzhen},
  journal={IEEE Journal of Selected Topics in Applied Earth Observations and Remote Sensing}, 
  title={Ground Clutter and Noise Mitigation Based on Range–Doppler Spectral Processing for Polarimetric Weather Radar}, 
  year={2024},
  volume={17},
  number={},
  pages={1026-1045},
  doi={10.1109/JSTARS.2024.3420074   }}

@ARTICLE{li2014ground,
  author={Li, Yinguang and Zhang, Guifu and Doviak, Richard J.},
  journal={IEEE Transactions on Signal Processing}, 
  title={Ground Clutter Detection Using the Statistical Properties of Signals Received With a Polarimetric Radar}, 
  year={2014},
  volume={62},
  number={3},
  pages={597-606},
  doi={10.1109/TSP.2013.2293118  }}

@ARTICLE{an2024radio,
  author={An, Mengyun and Yin, Jiapeng and Liu, Ting and Wu, Zezhou and Li, Yongzhen},
  journal={IEEE Transactions on Geoscience and Remote Sensing}, 
  title={Radio Frequency Interference Mitigation Based on Low-Rank Sparse Decomposition for Polarimetric Weather Radar}, 
  year={2024},
  volume={62},
  number={},
  pages={1-19},
  doi={10.1109/TGRS.2024.3414302  }}

@ARTICLE{yin2022radio,
  author={Yin, Jiapeng and Hoogeboom, Peter and Unal, Christine and Russchenberg, Herman},
  journal={IEEE Transactions on Geoscience and Remote Sensing}, 
  title={Radio Frequency Interference Characterization and Mitigation for Polarimetric Weather Radar: A Study Case}, 
  year={2022},
  volume={60},
  number={},
  pages={1-16},
  doi={10.1109/TGRS.2021.3093565 }}

@ARTICLE{lin2017multi,
  author={Lin, Wei and Wong, Hang},
  journal={IEEE Antennas and Wireless Propagation Letters}, 
  title={Multipolarization-Reconfigurable Circular Patch Antenna With L-Shaped Probes}, 
  year={2017},
  volume={16},
  number={},
  pages={1549-1552},
  doi={10.1109/LAWP.2017.2648862}}

@ARTICLE{nishamol2011an,
  author={Nishamol, M. S. and Sarin, V. P. and Tony, D. and Aanandan, C. K. and Mohanan, P. and Vasudevan, K.},
  journal={IEEE Transactions on Antennas and Propagation}, 
  title={An Electronically Reconfigurable Microstrip Antenna With Switchable Slots for Polarization Diversity}, 
  year={2011},
  volume={59},
  number={9},
  pages={3424-3427},
  doi={10.1109/TAP.2011.2161446}}

@ARTICLE{chen2021a,
  author={Chen, Dingzhao and Liu, Yanhui and Chen, Shu-Lin and Qin, Pei-Yuan and Guo, Y. Jay},
  journal={IEEE Transactions on Antennas and Propagation}, 
  title={A Wideband High-Gain Multilinear Polarization Reconfigurable Antenna}, 
  year={2021},
  volume={69},
  number={7},
  pages={4136-4141},
  doi={10.1109/TAP.2020.3044395}}

@ARTICLE{yang2021recon,
  author={Yang, Zhengyi and Kou, Na and Yu, Shixing and Long, Fei and Yuan, Lili and Ding, Zhao and Zhang, Zhengping},
  journal={IEEE Microwave and Wireless Components Letters}, 
  title={Reconfigurable Multifunction Polarization Converter Integrated With PIN Diode}, 
  year={2021},
  volume={31},
  number={6},
  pages={557-560},
  doi={10.1109/LMWC.2021.3064039}}

@ARTICLE{yang2023a,
  author={Yang, Heng and Wang, Shi Cong and Li, Peng and He, Yuan and Zhang, Yun Jing},
  journal={IEEE Transactions on Antennas and Propagation}, 
  title={A Broadband Multifunctional Reconfigurable Polarization Conversion Metasurface}, 
  year={2023},
  volume={71},
  number={7},
  pages={5759-5767},
  doi={10.1109/TAP.2023.3266498}}

@ARTICLE{ludwig1973the,
  author={Ludwig, A.},
  journal={IEEE Transactions on Antennas and Propagation},
  title={The definition of cross polarization},
  year={1973},
  volume={21},
  number={1},
  pages={116-119},
  doi={10.1109/TAP.1973.1140406}}

@ARTICLE{pozar1994the,
  author={Pozar, D.M.},
  journal={IEEE Transactions on Antennas and Propagation}, 
  title={The active element pattern}, 
  year={1994},
  volume={42},
  number={8},
  pages={1176-1178},
  doi={10.1109/8.310010}}

@ARTICLE{xie2025coarray,
  author={Xie, Qianpeng and Wang, Zhanling and Wen, Fangqing and He, Jin and Truong, Trieu-Kien},
  journal={IEEE Transactions on Antennas and Propagation}, 
  title={Coarray Tensor Train Decomposition for Bistatic MIMO Radar With Uniform Planar Array}, 
  year={2025},
  volume={73},
  number={8},
  pages={5310-5323},
  doi={10.1109/TAP.2025.3567446}}

\begin{IEEEbiography}[{\includegraphics[width=1in,height=1.25in,clip,keepaspectratio]{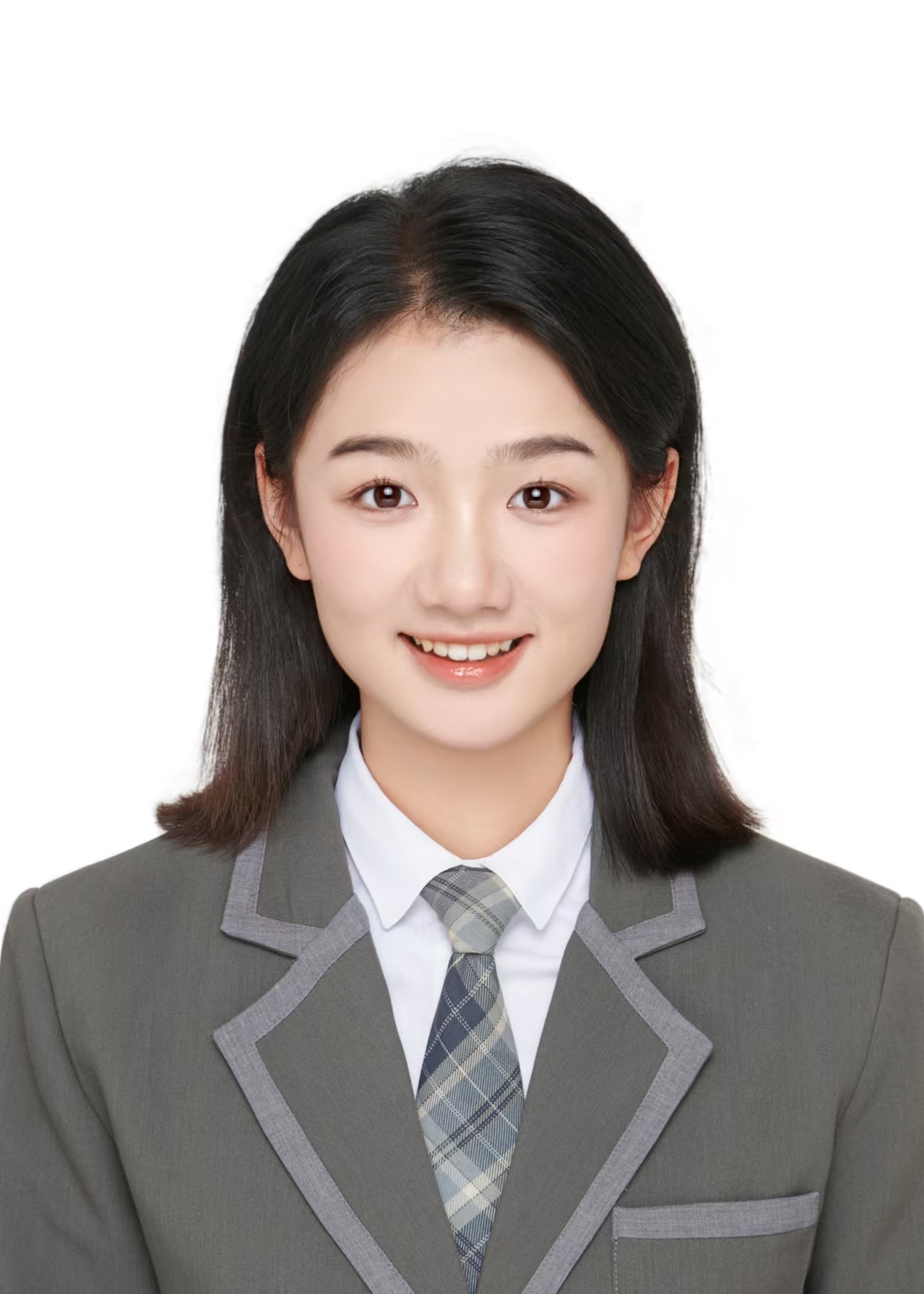}}]{Yiqing Wang}
was born in 1999. She received the B.S. degree in information engineering from the National University of Defense Technology, Changsha, China, in 2021, where she is currently pursuing the Ph.D. degree in information and communication engineering. Her research interests include antenna arrays and polarimetric phased arrays.
\end{IEEEbiography}

\begin{IEEEbiography}[{\includegraphics[width=1in,height=1.25in,clip,keepaspectratio]{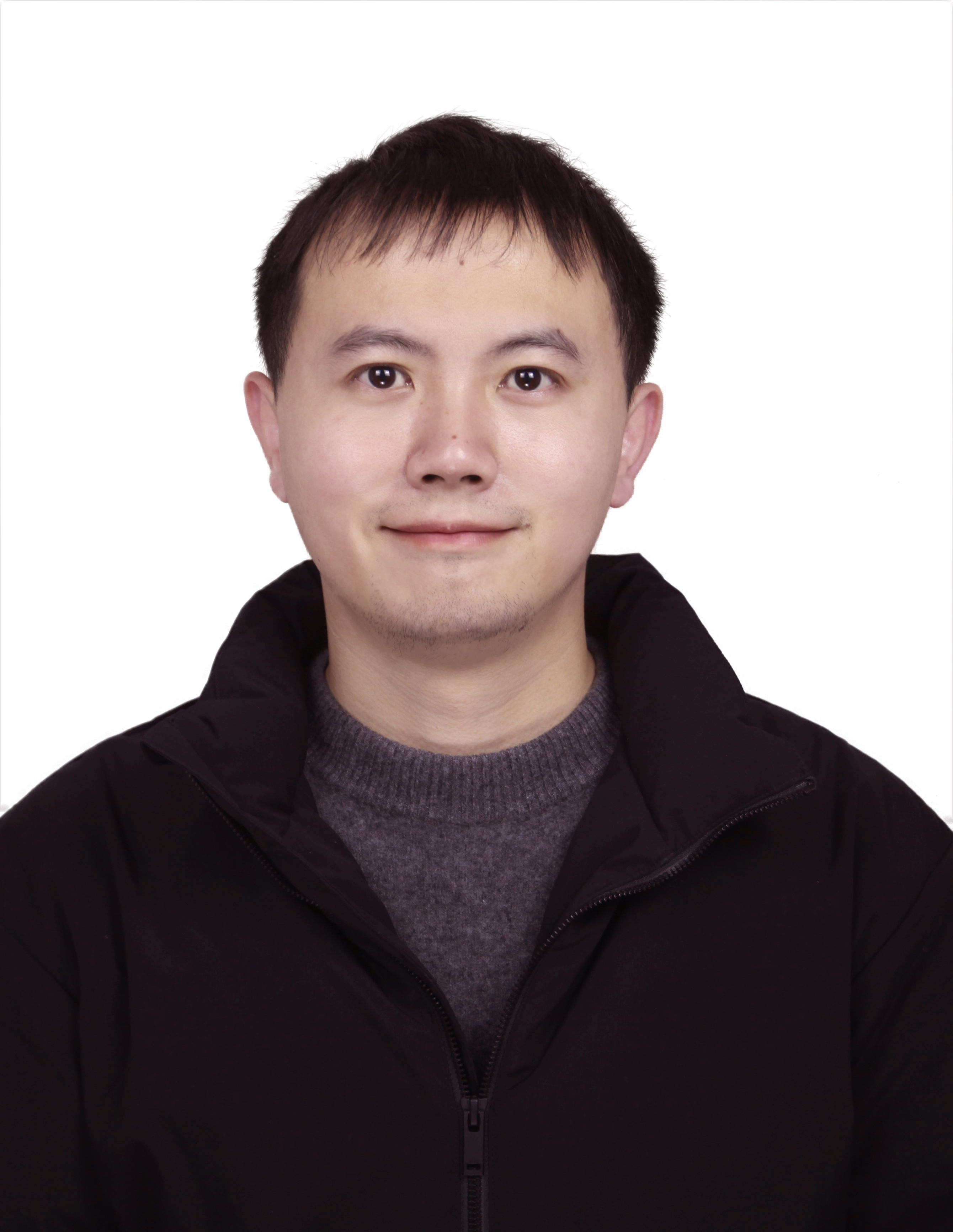}}]{Jian Zhou}
was born in 1996. He received the B.S. degree in information engineering and the Ph.D. degree in information and communication engineering from the National University of Defense Technology (NUDT), Changsha, China, in 2018 and 2024, respectively. In 2025, he joined the College of Electronic Science, NUDT, where he is currently an Assistant Researcher. His research interests include antenna arrays and polarimetric phased array.
\end{IEEEbiography}

\begin{IEEEbiography}[{\includegraphics[width=1in,height=1.25in,clip,keepaspectratio]{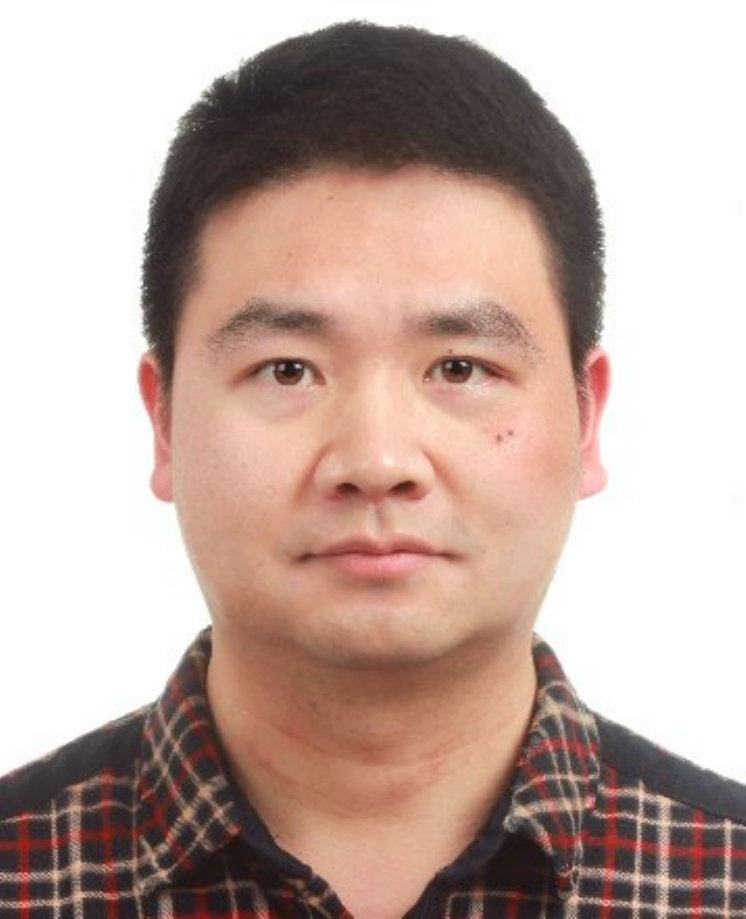}}]{Chen Pang}
was born in 1986. He received the B.S. and Ph.D. degrees in information and communication engineering from the National University of Defense Technology (NUDT), Changsha, China, in 2009 and 2015, respectively. From November 2012 to November 2014, he was a visiting Ph.D. Student with the Delft University of Technology, Delft, The Netherlands. 
\par Currently, he is an Associate Professor with the College of Electronic Science and Technology, National University of Defense Technology. His research interests include weather radar signal
processing, radar polarimetry, active array, and electromagnetic simulation.
\end{IEEEbiography}

\begin{IEEEbiography}[{\includegraphics[width=1in,height=1.25in,clip,keepaspectratio]{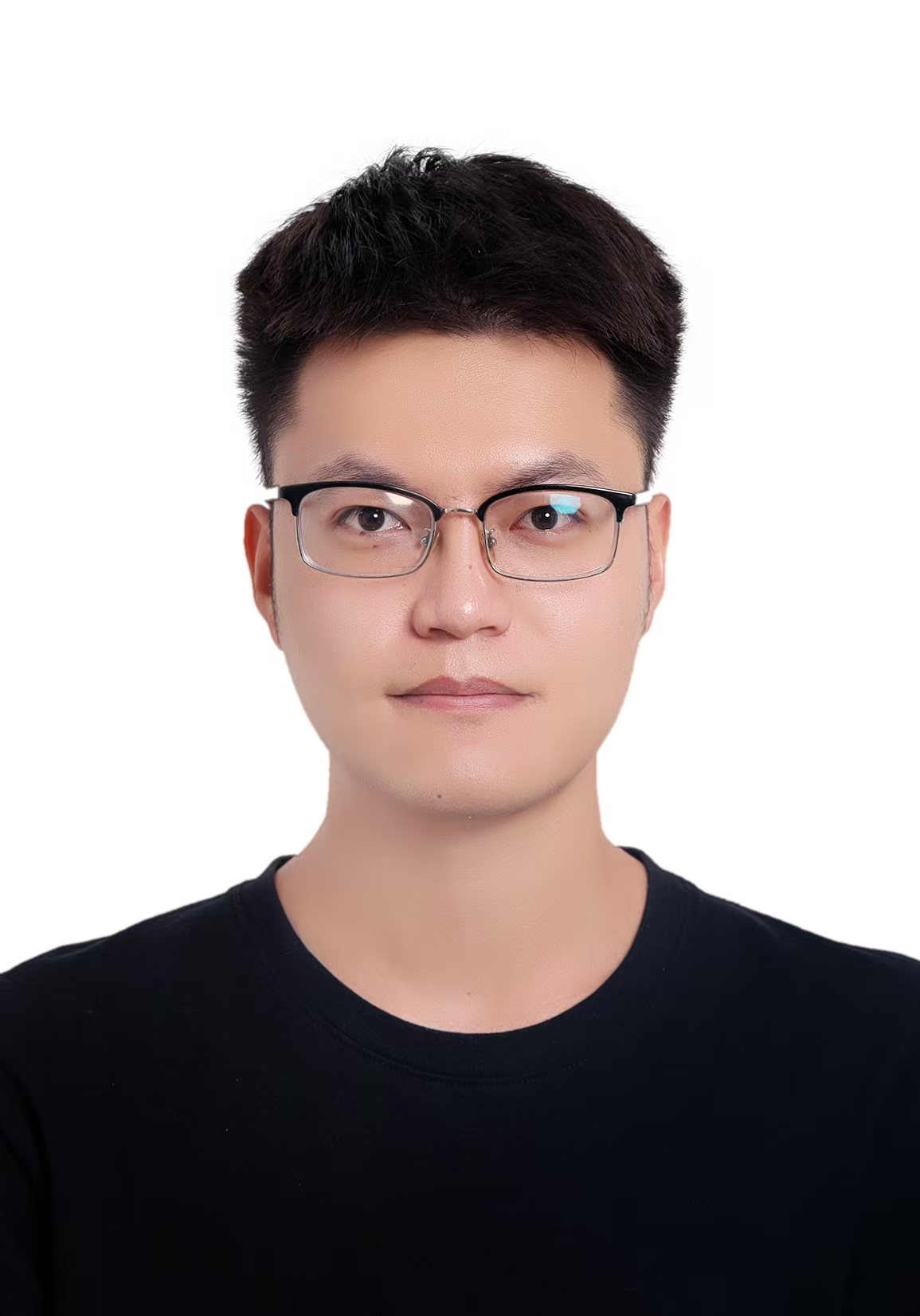}}]{Wenyang Man}
was born in 2000. He received the B.S. degree in Electronic Engineering from North China Electric Power University, Baoding, China, in 2023, and is currently pursuing the Ph.D. degree in Electronic Science and Technology at the National University of Defense Technology, Changsha, China. His research interests include shared-aperture polarimetric phased array radar.
\end{IEEEbiography}

\begin{IEEEbiography}[{\includegraphics[width=1in,height=1.25in,clip,keepaspectratio]{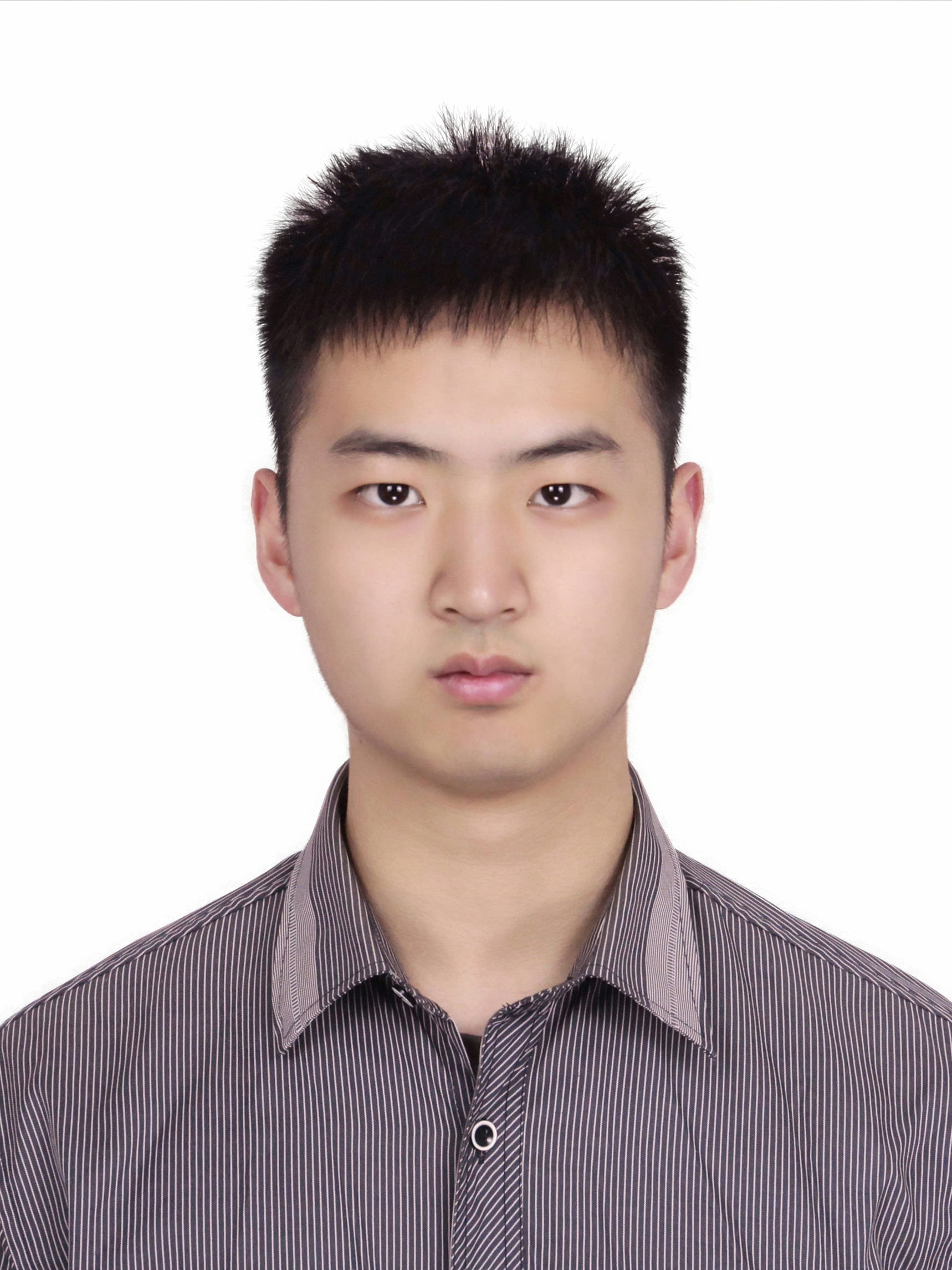}}]{Zixiang Xiong}
was born in 2000. He received the M.S. degree in information engineering from the National University of Defense Technology, Changsha, China, in 2025, where he is currently pursuing the Ph.D. degree in electronic information. His research interests include antenna arrays and polarimetric phased arrays.
\end{IEEEbiography}

\vspace{-50 mm}

\begin{IEEEbiography}[{\includegraphics[width=1in,height=1.25in,clip,keepaspectratio]{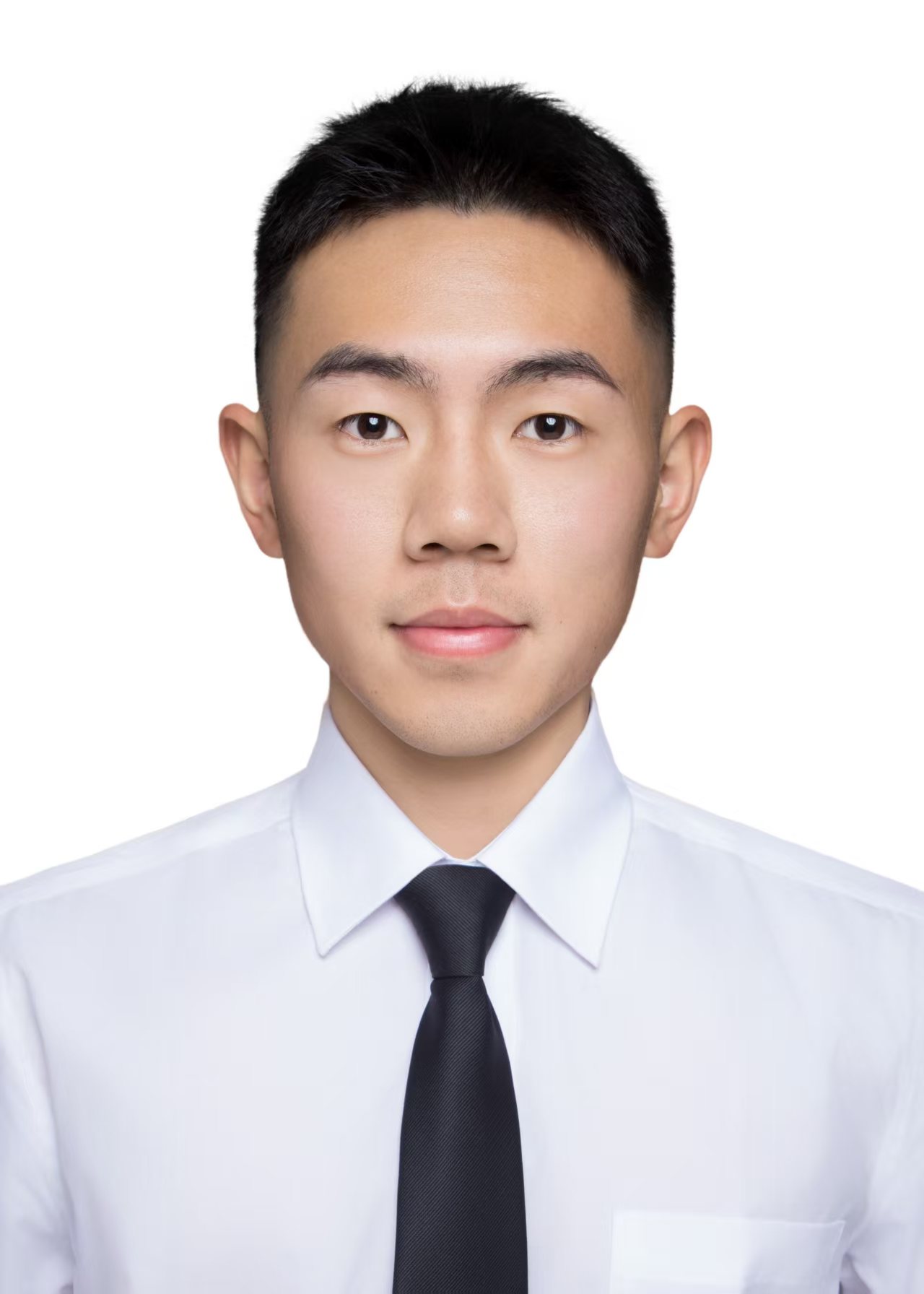}}]{Ke Meng}
was born in 1999. He received the B.S. degree in information engineering from the National University of Defense Technology, Changsha, China, in 2021, where he is currently pursuing the M.S. degree in electronic information. His research interests include antenna arrays and polarimetric phased arrays.
\end{IEEEbiography}

\vspace{-50 mm}

\begin{IEEEbiography}[{\includegraphics[width=1in,height=1.25in,clip,keepaspectratio]{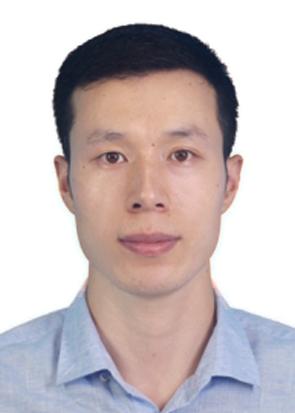}}]{Zhanling Wang}
was born in 1988. He received the B.S. degree in electronic engineering from Shenyang University of Technology, Shenyang, China, in 2007, the M.S. degree in information and communication engineering from Air Force Engineering University,
Xi’an, China, in 2011, and the Ph.D. degree from the National University of Defense Technology, Changsha, China, in 2021.
\par He is currently a Lecturer. His research interests include array antennas design and polarimetric phased arrays.
\end{IEEEbiography}

\vspace{-50 mm}

\begin{IEEEbiography}[{\includegraphics[width=1in,height=1.25in,clip,keepaspectratio]{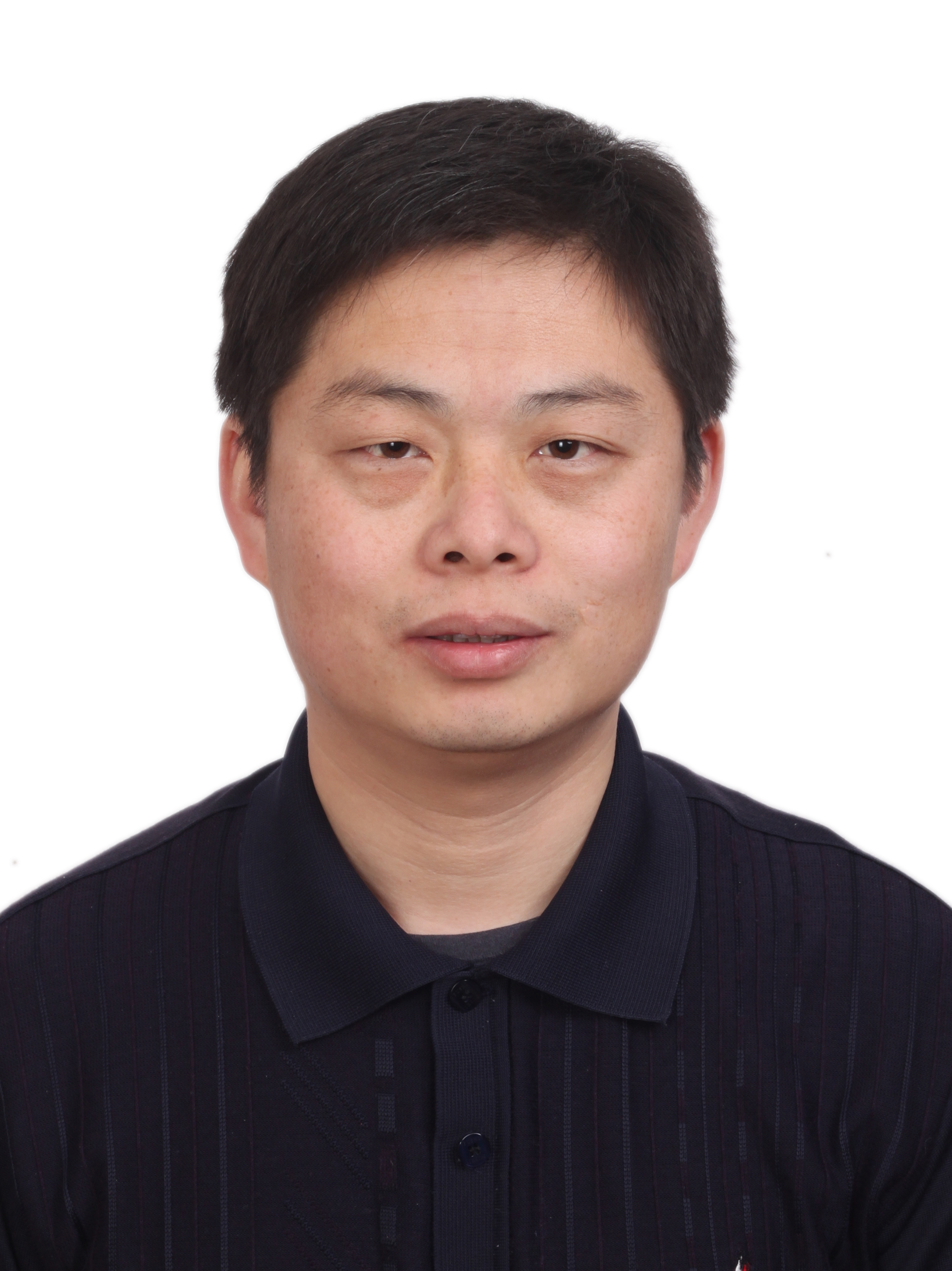}}]{Yongzhen Li}
was born in 1977. He received the B.S. and Ph.D. degrees in electronic engineering from the National University of Defense Technology, Changsha, China, in 1999 and 2004, respectively. 
\par Currently, he is a Professor with the National University of Defense Technology. His research interests include radar signal processing, radar polarimetry, and target recognition.
\end{IEEEbiography}

\end{document}